\documentclass[sigconf]{acmart}

\AtBeginDocument{%
  }

\setcopyright{none}
\renewcommand\footnotetextcopyrightpermission[1]{}

\usepackage{tikz}
\usepackage{amsmath}
\usepackage{threeparttable}
\usepackage{graphicx}
\usepackage[many]{tcolorbox}
\usepackage{mathtools,latexsym,amsfonts,stmaryrd}

\usepackage{booktabs} % added for tables
\usepackage{pbox}
\usepackage{amsthm}
\usepackage{amsmath}
\usepackage{multirow}
\usepackage{tikz}
\usepackage{mathrsfs}
\usepackage{mathpartir}
\usepackage[ruled,linesnumbered,vlined,noend]{algorithm2e}
\usepackage{url}
\usepackage{syntax}
\usepackage{framed}
\usepackage[noend]{algpseudocode}
\usepackage{flushend}
\usepackage{listings}
\usepackage{moresize}
\usepackage{wrapfig}
 
\usepackage{seqsplit}
\usepackage{alltt}
\usepackage{xspace}
\usepackage[normalem]{ulem}
\usepackage{enumitem}
\usepackage{pifont}% http://ctan.org/pkg/pifont
\DeclareMathAlphabet{\mathcal}{OMS}{cmsy}{m}{n}

\usepackage{amsmath, listings, amsthm, proof, xspace}

\usepackage{thmtools}

\declaretheoremstyle[spaceabove=\topsep,notefont=\normalfont\itshape]{mystyle}

\definecolor{ForestGreen}{RGB}{34,139,34}

\newcommand{\revise}[2]{{\color{red}{\ifx&#1&\else- #1\fi}} {\color{ForestGreen}{\ifx&#2&\else+ #2\fi}}}%
\renewcommand{\revise}[2]{#2}%

\usetikzlibrary{arrows,patterns, decorations.pathreplacing}

\newcommand{\finding}[2]{
  \smallskip
  \smallskip
\begin{tcolorbox}[width=\linewidth,boxrule=0pt,top=1pt, bottom=1pt, left=1pt,right=1pt, colback=gray!20,colframe=gray!20]
\textbf{Finding #1:} %\textit
{#2}
\end{tcolorbox}}

\newcommand{\F}{Figure\xspace}

\newcommand{\T}{Table}
\newcommand{\mysec}{\S}

\newcommand{\ignore}[1]{}
\newcommand{\mysubref}[2]{\hyperref[#1]{\ref*{#1}(#2)}}
\newcommand{\AP}{Appendix}

\lstdefinestyle{base}{
  moredelim=**[is][\color{red}]{@}{@},
  escapeinside={<@}{@>}
}

\usepackage{pifont}% http://ctan.org/pkg/pifont

\newcommand\DejaVuttfamily{%
  \fontfamily{DejaVuSansMono-TLF}\selectfont
}

\lstdefinestyle{base}{
  moredelim=**[is][\color{red}]{@}{@},
  escapeinside={<@}{@>}
}

\lstdefinelanguage
   [x64]{Assembler}     % add a "x64" dialect of Assembler
   [x86masm]{Assembler} % based on the "x86masm" dialect
   {morekeywords={CDQE,CQO,CMPSQ,CMPXCHG16B,JRCXZ,LODSQ,MOVSXD, %
                  POPFQ,PUSHFQ,SCASQ,STOSQ,IRETQ,RDTSCP,SWAPGS, %
                  rax,rdx,rcx,rbx,rsi,rdi,rsp,rbp, %
                  r8,r8d,r8w,r8b,r9,r9d,r9w,r9b,reg128,m128}} % etc.

\let\OLDthebibliography\thebibliography
\renewcommand\thebibliography[1]{
  \OLDthebibliography{#1}
  \setlength{\parskip}{0pt}
  \setlength{\itemsep}{1pt plus 0.85ex}
}
\usepackage{listings}
\usepackage{fancyvrb}
\usepackage{color}
\definecolor{lightgray}{rgb}{.9,.9,.9}
\definecolor{darkgray}{rgb}{.4,.4,.4}
\definecolor{purple}{rgb}{0.65, 0.12, 0.82}
\definecolor{commentgreen}{RGB}{63,127,95}
\definecolor{pyblue}{RGB}{59,117,175}
\definecolor{pyorange}{RGB}{239,134,54}
\definecolor{pygreen}{RGB}{81,158,62}

\usepackage{xcolor}
\colorlet{myPurple}{blue!40!red}
\definecolor{myOrange}{RGB}{255,192,0}

\lstdefinelanguage{Solidity}{
  keywords={len,delete,int,void,payable, public, event, contract, typeof, new, true, false, catch, function, return, null, catch, switch, var, if, while, do, else, case, break,struct,const,socklen_t,sa_familty_t,char,sockaddr,load},
  keywordstyle=\color{violet}\bfseries,
  ndkeywords={class, export, boolean, throw, implements, import, this},
  ndkeywordstyle=\color{darkgray}\bfseries,
  identifierstyle=\color{black},
  sensitive=false,
  comment=[l]{//},
  escapeinside={(*@}{@*)},          % if you want to add LaTeX within your code
  morecomment=[s]{/*}{*/},
  commentstyle=\color{commentgreen}\ttfamily,
  stringstyle=\color{red}\ttfamily,
  morestring=[b]',
  morestring=[b]"
}

\newcommand{\rnum}[1]{\uppercase\expandafter{\romannumeral #1\relax}}

\usepackage{xcolor}% http://ctan.org/pkg/xcolor

\algnewcommand{\LeftComment}[1]{\Statex \(\triangleright\) #1}

\definecolor{pptbrown}{RGB}{132,60,12}
\definecolor{pptgreen}{RGB}{56,87,35}
\definecolor{pptred}{RGB}{155,30,20}
\definecolor{pptdy}{RGB}{127,96,0}
\definecolor{matplotlib-cyan}{HTML}{17becf}

\newcommand{\parh}[1]{\noindent\textbf{#1}}

\newcommand{\tool}{\textsc{DDE}\xspace}
\newcommand{\dsr}{\textsc{DSR}\xspace}
\newcommand{\va}{\textsc{Vanilla}\xspace}

\newcommand{\threshold}{$\tau$\xspace}
\newcommand{\maxbranch}{\textsc{MBR}\xspace}

\newcommand{\rom}[1]{\uppercase\expandafter{\romannumeral #1\relax}}

\newcommand{\CBrush}{\textcolor[RGB]{84,130,53}{\ding{51}}}
\newcommand{\XBrush}{\textcolor[RGB]{176,35,24}{\ding{55}}}

\newcommand{\mybox}[1]{
\begin{tcolorbox}[boxrule=0pt,frame hidden,sharp corners,enhanced,borderline west={2pt}{0pt}{black}]
#1
\end{tcolorbox}
}

\theoremstyle{definition}
\newtheorem{exmp}{Example}[section]

\begin{document}

\title{Differentiation-Based Extraction of Proprietary Data from Fine-Tuned LLMs}

\author{Zongjie Li}
\email{zligo@cse.ust.hk}
\orcid{0000-0002-9897-4086}
\affiliation{
 \institution{Hong Kong University of Science and Technology}
 \city{Hong Kong}
 \country{China}
}

\author{Daoyuan Wu}
\authornotemark[1] 
\email{daoyuan@cse.ust.hk}
\orcid{0000-0002-3752-0718}
\affiliation{
 \institution{Hong Kong University of Science and Technology}
 \city{Hong Kong}
 \country{China}
}

\author{Shuai Wang}
\authornote{Corresponding authors.} 
\email{shuaiw@cse.ust.hk}
\orcid{0000-0002-0866-0308}
\affiliation{
 \institution{Hong Kong University of Science and Technology}
 \city{Hong Kong}
 \country{China}
}

\author{Zhendong Su}
\email{zhendong.su@inf.ethz.ch}
\orcid{0000-0002-2970-1391}
\affiliation{%
\institution{ETH Zurich}
\city{Zurich}
\country{Switzerland}}

\begin{abstract}

The increasing demand for domain-specific and human-aligned Large Language Models (LLMs) has led to the widespread adoption of Supervised Fine-Tuning (SFT) techniques.
SFT datasets often comprise valuable instruction-response pairs,
making them highly valuable targets for potential extraction.
This paper studies this critical research problem for the first time. 
We start by formally defining and formulating the problem, then explore various attack goals, types, and variants based on the unique properties of SFT data in real-world scenarios.
Based on our analysis of extraction behaviors of direct extraction, we develop a novel extraction method specifically designed for SFT models, called Differentiated Data Extraction (\tool), which exploits the confidence levels of fine-tuned models 
and their behavioral differences from pre-trained base models.  
Through extensive experiments across multiple domains and scenarios, we demonstrate the feasibility of SFT data extraction using \tool.
Our results show that \tool consistently outperforms existing extraction baselines in all attack settings.
To counter this new attack, we propose a defense mechanism that mitigates \tool attacks with minimal impact on model performance.
Overall, our research reveals hidden data leak risks in fine-tuned LLMs and provides insights for developing more secure models.
\end{abstract}

\maketitle

\section{Introduction}
\label{sec:introduction}

The rapid advancement of Large Language Models (LLMs) has led to remarkable achievements, with models like GPT-3~\cite{brown2020language} and PaLM~\cite{chowdhery2022palm} demonstrating human-level performance across various tasks~\cite{brown2020language,li2025api}.
These models are extensively pre-trained on vast and diverse datasets, including web pages, books, and academic articles, which enables them to acquire a broad range of knowledge and capabilities.
However, despite their impressive scale in terms of data and parameters, LLMs still face significant challenges in specialized contexts.
The increasing demands for these models to excel in domain-specific tasks and to align with human expectations underscores limitations that hinder their widespread adoption.

To enhance LLMs, researchers employ Supervised Fine-Tuning (SFT) (detailed in \mysec\ref{sec:preliminary}) as a post-training solution. This approach utilizes SFT datasets, consisting of instruction ($I$) and response ($R$) pairs, to encapsulate task-specific knowledge. Unlike vast pre-training datasets, SFT data is more valuable, significantly smaller, and used differently in training. These datasets, typically curated through manual effort or refinement of existing high-quality data, are used to fine-tune pre-trained models. The resulting models, denoted as $M_{FT}$, demonstrate improved performance in specific domains and better alignment with human expectations, making them more suitable for targeted applications.

Given that $M_{FT}$ inherently contains knowledge from the SFT data, a natural question arises:
\textit{Is it possible to extract the SFT data from an fine-tuned LLM?}
While data extraction techniques such as DSR~\cite{zanella2020analyzing}
have been studied for traditional machine learning
models~\cite{salem2020updates,jagielski2023combine,hui2023information},
extracting SFT data from LLMs presents inherently different challenges. These
differences stem from both LLMs' unique training paradigm and their structured
I-R pairs (detailed in \mysec\ref{sec:preliminary}). Moreover, prior studies on
LLMs have focused on the extraction of pre-training
data~\cite{carlini2021extracting,nasr2023scalable}. However, these works
primarily aim to extract and verify a subset of the data potentially used in the
pre-training process. This is due to the enormous volume of pre-training data
and its lack of explicit ($I$-$R$) pairs, as pre-training data typically
consists of unstructured text corpora~\cite{2019t5}. In contrast, SFT data
extraction presents unique challenges and implications. The strict
correspondence between $I$-$R$ pairs, along with the potential for attackers to
use extracted SFT data for their own fine-tuning purposes to replicate the
victim model's functionality, renders existing methods inadequate, as further
discussed in \mysec\ref{subsec:attack-types-comparison}. Consequently, there
exists a significant research gap in SFT data extraction.

In this paper, we study the problem of SFT data extraction for the first time. We begin by formally defining and formulating the problem, introducing two distinct attack goals based on attacker objectives: \textit{reconstruction} and \textit{retraining}.
The former aims to accurately recover the original SFT data (e.g., valuable domain data on Alzheimer's disease diagnosis) with high fidelity, while the latter seeks to use the extracted data for further fine-tuning models to achieve comparable capabilities.
Furthermore, considering the unique properties of $I$-$R$ pairs, we propose two attack types (R-I and I-R attack), where I-R aims to extract responses from instructions and R-I vice versa. To account for real-world attack scenarios, we introduce three possible attack variants based on different instruction preservation methods. Through these various attack goals, types, and variants, we aim to conduct a thorough and detailed feasibility study of SFT data extraction.

Building upon the feasibility study, we conduct a pilot investigation
revealing the limitations of the direct extraction approach (referred to
as the \va approach), which often fails to extract SFT data due to errors
propagating from earlier positions in the extracted sequence. To address this,  we propose a novel and effective attack called
\textit{Differentiate Data Extraction} (\tool). \tool leverages two key insights:
(1) the fine-tuned model's higher confidence in generating SFT data, and (2)
behavioral differences between fine-tuned and base models. 
By identifying potential ``branch deviation points'' and comparing
generation branches from both models, \tool selects sequences more likely to
reflect true SFT data.

Our comprehensive experiments demonstrate the feasibility of SFT data extraction across various attack goals and scenarios. We find that preservation methods significantly influence attack performance, with higher retention rates generally yielding superior results. Substantial variations in effectiveness across SFT domains and attack types underscore the attack's complexity, while retraining attacks show greater resilience to different preservation methods compared to reconstruction attacks. 
Our analysis reveals \tool's consistent superiority over both
\va and \dsr baselines in reconstruction attacks, with average
improvements of 9.96\% and 5.73\%, respectively. For retraining attacks, \tool
demonstrates strong performance with improvements of 9.41\% over \va\ and
11.52\% over \dsr. These gains highlight \tool's enhanced ability to accurately
extract SFT data and enable competitive model performance compared to victim
models. Further exploration provides insights into key factors affecting \tool's
performance, including base model selection and hyperparameter tuning.
Additionally, we propose a defense method, which can be used to fail \tool on
the extracted data while influencing the model performance within 3\%.
In summary, our contributions are as follows:
\begin{itemize}
    \item We present the first comprehensive study on the feasibility of extracting SFT data from fine-tuned LLMs, addressing growing privacy concerns. We formulate the SFT data extraction problem by introducing distinct attack goals, types, and variants, providing a structured approach to evaluate the vulnerability of fine-tuned LLMs.

    \item We propose Differentiate Data Extraction (\tool), a novel method that leverages model confidence and behavioral differences between fine-tuned and base models for improved extraction accuracy.

    \item Through extensive experiments across various SFT domains and attack
    scenarios, we demonstrate the feasibility of SFT data extraction and
    reveal distinct characteristics between reconstruction and
    retraining attacks, clarifying the necessity of seperately handling them.
    Our results show that \tool consistently outperforms existing baselines across both attack goals.

    \item We propose a defense mechanism that can effectively mitigate \tool
    attacks while minimally impacting the model's performance, offering a
    practical solution to enhance the privacy of fine-tuned LLMs.
\end{itemize}

\section{Preliminary}
\label{sec:preliminary}

\begin{figure}[!tbp]
    \centering
    \includegraphics[width=0.9\linewidth]{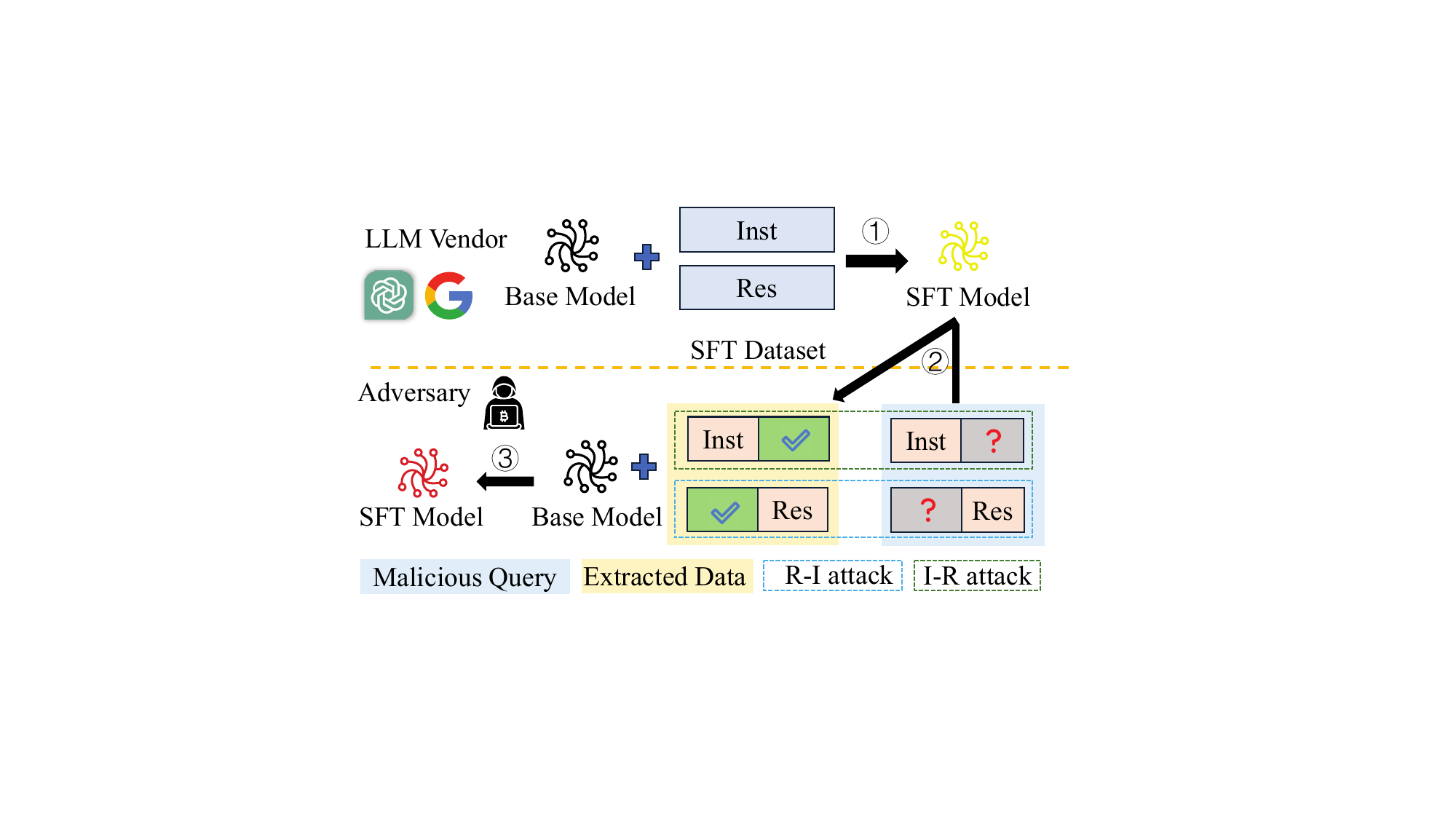}
    \caption{Attack scenario on SFT models. ``Inst'' and ``Res'' are instructions and responses, respectively. \ding{172} LLM vendor tunes an SFT model from a base model using an SFT dataset. \ding{173} Adversary extracts data from the SFT model via malicious queries (I-R or R-I attacks; see~\mysec\ref{sec:threat-model}). \ding{174} Adversary tunes a base model with extracted data to build their own SFT model.}
    \label{fig:attack-scenario}
\end{figure}

\parh{Supervised Fine-tuning.} 
Following the pre-training phase, LLMs typically undergo additional fine-tuning steps to better align with human intentions and task-specific requirements.
Among these refinement techniques, supervised fine-tuning (SFT) has emerged as a prevalent and effective approach~\cite{chung2024scaling,brown2020language}.
The SFT process, as depicted in \F~\ref{fig:attack-scenario} step \ding{172}, utilizes SFT datasets composed of instruction-response pairs $\{(i^d, r^d)\}$, where $i^d$ denotes the input instruction and $r^d$ represents the corresponding desired response.
This structured format is distinct from pre-training data, enabling the model to acquire task-specific behaviors and significantly improve its instruction-following capabilities.
Formally, for a specific domain $d$ with context $c^d$, the SFT objective is to minimize the negative log-likelihood of the target response $r^d$ given the context $c^d$ and input instruction $i^d$. That is:
\begin{equation}
L_{\text{SFT}}(\theta) = -\log f_\theta(r^d | c^d, i^d),
\end{equation}
where $\theta$ represents the model parameters. Upon completion of the SFT
process, the resulting models are widely deployed to provide various
domain-specific online services, including but not limited to
medicine~\cite{singhal2023large} and finance~\cite{cheng2024adapting}. Notably,
in the context of this paper, we use SFT to only refer to full parameter
supervised fine-tuning. Those parameter-efficient fine-tuning methods~\cite{hu2021lora,liu2021p}
are not in our study scope because of their limited capabilities in
understanding the knowledge; see more discussion in~\AP~\ref{app:peft}.

\parh{Data Extraction Attack.}
Data extraction attacks aim to recover training data from machine learning models. \T~\ref{tab:extraction-approaches} provides an overview of existing model extraction approaches. As shown in the table, early studies focus on traditional machine learning models with a relatively small number of parameters (\textless{}110M), targeting in image~\cite{salem2020updates,jagielski2023combine} and natural language processing~\cite{zanella2020analyzing,hui2023information} domains. 
These methods predominantly emphasize reconstruction attacks and are limited to extracting short text segments or structured data, making them inadequate for model retraining purposes and distinct from our approach.
Additionally, the training paradigm of these works remain consistent during model updates, with unchanged input-output formats and loss functions. In contrast, our work addresses the unique challenges of SFT
data extraction, where the training paradigm shifts significantly from
pre-training to fine-tuning stages, introducing new training objectives and data formats.

For LLMs, current research~\cite{carlini2021extracting,nasr2023scalable} mainly focus on extracting pre-training data from large-scale models (7B+ parameters), which typically involves comparing model responses against large-scale web-scraped datasets. As indicated in \T~\ref{tab:extraction-approaches}, these attacks are inherently untargeted due to the vast scale of pre-training datasets. 
These untargeted methods are insufficient for SFT data extraction where precise instruction-response pairing is crucial, a distinction we further elaborate in \S\ref{subsec:attack-types-comparison} with experimental evidence.
In contrast, we focus on the targeted extraction of specific, high-value information used in the fine-tuning process. We use malicious queries in two formats: instruction-to-response (I-R) and response-to-instruction (R-I), as shown in \F~\ref{fig:attack-scenario} step \ding{173}. These extracted SFT data enable attackers to further fine-tune their models (\F~\ref{fig:attack-scenario} step \ding{174}), potentially replicating the victim model's capabilities.

\begin{table*}[!t]
    \caption{Overview of various model extraction approaches, with most targeting pre-trained models and only \tool targeting SFT models. \CBrush, \XBrush\ denote whether a method supports a specific capability. ``Long'' indicates whether the method supports extracting long text segments. ``Size'' denotes the maximum number of parameters of victim models to which the method is applied.}
    \centering
    \resizebox{0.95\linewidth}{!}{
    \begin{tabular}{lcccccccc}
    \hline
    Method & Model Type & Size & Attack Type & Reconstruction & Retraining & Long & Object & Task \\ \hline
    Hui et al.~\cite{hui2023information} & Pretrain & \textless{}10M & Targeted & \CBrush & \XBrush & \XBrush & Record & Classification \\
    Jagielski et al.~\cite{jagielski2023combine} & Pretrain & \textless{}110M & Targeted & \CBrush & \XBrush & \XBrush & Image/Record & Classification \\
    Salem et al.~\cite{salem2020updates} & Pretrain & \textless{}75M & Targeted & \CBrush & \XBrush & \XBrush & Image & Classification \\ \hline
    Poem~\cite{nasr2023scalable}/Random~\cite{carlini2021extracting} & Pretrain & 7B+ & Untargeted & \CBrush & \CBrush & \CBrush & Text & Generation \\ 
    \dsr~\cite{zanella2020analyzing} & Pretrain & \textless{}15M & Targeted & \CBrush & \XBrush & \XBrush & Text & Generation \\
    \va\ (implemented by us) & Pretrain & 7B+ & Targeted & \CBrush & \CBrush & \CBrush & Text & Generation \\
    \tool (our method) & SFT & 7B+ & Targeted & \CBrush & \CBrush & \CBrush & Text & Generation \\ \hline
    \end{tabular}
    }
    \label{tab:extraction-approaches}
\end{table*}

\section{SFT Data Extraction Attack}
\label{sec:methodology}

This paper studies the SFT data extraction problem for the first time. In this
section, we start by formally defining and formulating the problem
(\mysec\ref{sec:threat-model}), and then explore multiple attack goals, types,
and variants
(\mysec\ref{sec:preservation-strategies}-\mysec\ref{subsec:attack-types-comparison}).
Lastly, in \mysec\ref{subsec:pilot-study}, we investigate the feasibility of
directly extracting SFT data. 
To ease reading, we summarize key symbols and categories used in this paper in \T~\ref{tab:notation}.

\subsection{Threat Model}
\label{sec:threat-model}

\parh{Scenario.}
We illustrate the attack scenario in \F~\ref{fig:attack-scenario}. Here, we
consider attackers who are users of an $M_{FT}$. Typically, LLM service vendors
employ SFT methods with specific datasets to fine-tune and get their models
($M_{FT}$), enhancing their ability to meet user requirements in real-world
applications. Exploiting the access to these fine-tuned models, the attackers
aim to extract the valuable SFT datasets used in the fine-tuning process. Their
objective is to recover this dataset by strategically querying the existing
$M_{FT}$. After acquiring the SFT dataset at low cost, attackers can further use
it to train their own models, enabling them to provide similar services.

\parh{Attacker's Capability.}~We assume that attackers cannot directly access
the backend LLM's weight information. They can only obtain the $M_{FT}$'s output
and corresponding token logits through queries. This assumption aligns with the
above attack scenario and real-world
solutions~\cite{apiopenai,togetherapi}, where LLM vendors usually 
provide the logits information for generated tokens. 
Specifically, we investigate 14 LLM service vendors, including both proprietary model vendors (e.g., OpenAI~\cite{apiopenai}) and open-source LLM API service vendors (e.g., Together AI~\cite{togetherapi}), to determine whether they provide logits information and assess the feasibility of our attack scenario. Our findings reveal that except for one platform that explicitly stated they do not support this feature, the remaining platforms either already provide or plan to provide such information. Details regarding the platform selection process can be found in~\AP~\ref{app:logits-availability}.
We assume that attackers can access a LLM with reasonable capacity (e.g.,
Gemma-7B~\cite{team2024gemma}, LLaMa2-7B~\cite{llama27b}), serving as $M_{Base}$. Although the
accessibility to $M_{FT}$'s corresponding $M_{Base}$ (i.e., $M_{FT}$ is
fine-tuned from $M_{Base}$) is not required, the accessibility of this ``real''
$M_{Base}$ enhances attacks (see \mysec\ref{subsec:ablation-study}). Also, due
to the unique nature of SFT data, we assume that the attacker has full or
partial knowledge of either the instruction ($I$) or the response ($R$).

\parh{Attacker's Type.}~Based on the attacker's knowledge of the targeted SFT
dataset, we define two types of attacks: I-R attack (known $I$, extracting $R$)
and R-I attack (known $R$, extracting $I$). Both types are significant in
practice as the value of $I$ or $R$ varies across domains. In the code domain,
well-crafted instructions ($I$) might be more valuable than code solutions
($R$), as specific guidelines meeting code requirements are rare and highly
valuable, making R-I attacks more critical. Conversely, in the medical domain,
expert diagnoses and treatments ($R$) are often more valuable than standard
symptom descriptions ($I$), as they represent sensitive medical knowledge,
rendering I-R attacks more consequential.

\parh{Attacker's Goal.}~For a dataset $D = \langle I, R \rangle$, attackers aim to recover the unknown component given partial information about the known component. They have two different objectives: \textit{reconstruction attack} and \textit{retraining attack}. The reconstruction attack focuses on the similarity between the extracted information and the original information, which is critical for restoring the original SFT data's information as much as possible, facilitating various downstream tasks such as copyright verification~\cite{duarte2024cop}. We formally express it with the I-R attack as:
\[ \min_{AM} \text{Dist}(AM(I'), R) \]
where $AM$ is the attack method, $I'$ is a variant of $I$ containing partial information (defined in~\mysec\ref{sec:preservation-strategies}), and Dist is a metric measuring the similarity between two strings (defined in~\mysec\ref{sec:setup}). The R-I attack follows a similar formulation with the roles of $I$ and $R$ reversed.

The retraining attack considers the effectiveness of the extracted SFT data in downstream applications. This is highly important as it allows attackers to obtain models with similar capabilities at a low cost. We define the effective rate (ER) for the
I-R attack as:
\[ \text{ER}_{I-R} = \text{Perf}(<I', AM(I')>,B) \]
where $B$ is the evaluation benchmark and $Perf$ is the performance of the SFT model fine-tuned with the corresponding extracted dataset.
ER for the R-I attack is defined analogously.
While one may argue that retraining does not inherently require data extraction, as techniques like distillation~\cite{hsieh2023distilling} or carefully curated query sets~\cite{zhou2024lima} can also produce high-performing models, we highlight a key distinction in our approach. Our retraining attacks specifically focus on extracting domain-specific knowledge from SFT datasets to enable models with similar capabilities in targeted domains, rather than merely optimizing for general benchmark performance. Furthermore, our approach assumes incomplete query knowledge (\S\ref{sec:preservation-strategies}), reflecting realistic attack scenarios where attackers have limited information. This differs significantly from existing methods that require complete access to queries.

\begin{table}[!tbp]
  \centering
  \caption{Key symbols and categories used in the paper.}
  \label{tab:notation}
  \resizebox{0.8\linewidth}{!}{

  \setlength{\tabcolsep}{3pt}
  \begin{tabular}{c|l}
      \hline
      \multicolumn{2}{c}{\textbf{Symbols and Abbreviations}} \\
      \hline
      \textbf{Notation} & \textbf{Description} \\
      \hline
      $M_{Base}$, $M_{FT}$ & Base and SFT models \\
      \threshold & Threshold \\
      \tool & Differentiate Data Extraction \\
      \hline
      \multicolumn{2}{c}{\textbf{Categories and Options}} \\
      \hline
      \textbf{Category} & \textbf{Options} \\
      \hline
      Attack goal & Reconstruction, Retraining \\
      Attack type & I-R, R-I \\
      Possible attack variants & PWP, PSP, SSP \\
      Attack method & \va, \dsr, \tool \\
      Retention rate & 25\%, 50\%, 75\% \\
      \hline
  \end{tabular}
  }
\end{table}

\subsection{Possible Attack Variants}
\label{sec:preservation-strategies}

\begin{figure}[!t]
    \centering
    \includegraphics[width=1.0\linewidth]{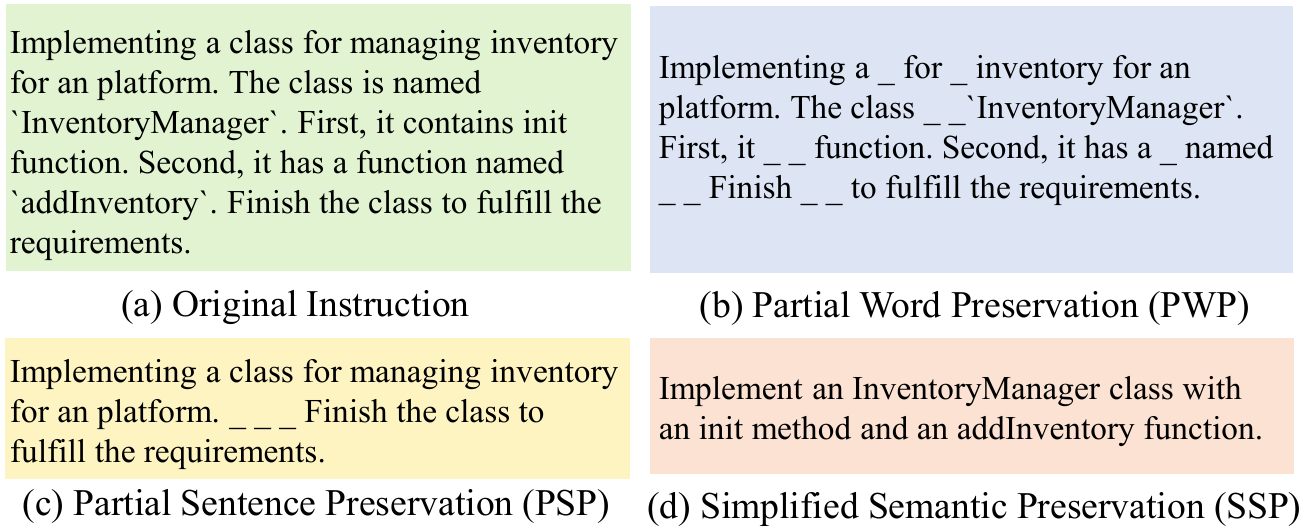}
    \caption{Simplified examples of possible attack variants using instruction preservation: (a) Original instruction, (b) Partial Word Preservation (PWP), (c) Partial Sentence Preservation (PSP), and (d) Simplified Semantic Preservation (SSP).}
    \label{fig:preservation}
\end{figure}

As discussed in~\mysec\ref{sec:threat-model}, we assume that attackers may only possess partial information of the I-R pair when executing their attacks.
This assumption is grounded in real-world scenarios where attackers can often infer the domain of target SFT models beforehand but are unlikely to have complete access to either component of the SFT data.
For example, attackers might construct plausible symptoms as potential responses for medical models when conducting R-I attacks, or formulate common math problems as potential instructions for math models in I-R attacks.
To comprehensively evaluate such realistic scenarios, we design three distinct preservation methods to accommodate possible attack variants. These methods not only reflect different levels of information accessibility but also simulate diverse real-world situations where partial data might be exposed.
Taking the I-R attack as an example, \F~\ref{fig:preservation} illustrates the possible attack variants, including the original instruction and our three proposed instruction preservation methods.

For all three methods, we define a retention rate (as noted in
\T~\ref{tab:notation}) to represent the portion of original information
accessible to attackers, considering rates of 25\%, 50\%, and 75\%.
We describe each method in detail below:

\parh{Partial Word Preservation (PWP).}~This scenario is inspired by users who share their LLM conversations with certain words redacted (e.g., due to privacy concerns). In this scenario, we assume the attacker has access to n\% of the words from the original instruction. To implement this approach, we randomly retain a portion of the words in the original instruction while masking the rest. We choose word-level preservation instead of token-level preservation because different LLMs may use different tokenizers, making it challenging to ensure consistency in the content that needs to be reconstructed.

\parh{Partial Sentence Preservation (PSP).}~This scenario is inspired by real-world instances of context leakage~\cite{leakageopenai}, where users might obtain partial chat contexts from other users, typically in the form of a few consecutive, meaningful sentences. Therefore, we assume the attacker in this case has access to n\% of the sentences from the original instruction. To implement this approach, we first segment the original instruction into sentences and then randomly retain a portion of these sentences while masking the rest.

\parh{Simplified Semantic Preservation (SSP).}~This scenario is inspired by the generalization of LLMs, which enables them to understand and respond similarly to semantically equivalent but differently expressed queries~\cite{yang2024exploring,vafa2024generalization}. Leveraging this characteristic, we assume the attacker in this case does not know the specific instruction but is aware of its semantic information or purpose.
To implement this approach, we use an LLM to rewrite the original instruction, preserving similar semantics but reducing the length to n\% of the original. We configure the rewrite instruction template as follows:

\mybox{Condense the following instruction to approximately \{n\}\% of its
original length without altering its core meaning. Preserve essential
information and intent:\{instruction\}. Provide only the revised instruction as
your response.}

Notably, these three preservation methods are not limited to I-R attacks; they can be equally applied to R-I attacks by simply inverting the roles of instructions and responses. For R-I attacks, the rewrite template would be adjusted accordingly to focus on the response rather than the instruction.

\subsection{Attack Types Comparison}
\label{subsec:attack-types-comparison}

\parh{I-R and R-I Attacks.}~As previously introduced, we categorize
attackers into two types: I-R and R-I. However, for the same \textless I,
R\textgreater\ pair, we posit that the R-I attack is significantly more
challenging than the I-R attack. 
To illustrate this point, let us consider the inference phase first. 
Given an LLM with an instruction, we can obtain a unique and deterministic response with greedy decoding, i.e., $R = LLM(I)$. However, this deterministic correspondence does not exist in the R-I scenario. For instance, when we know the response is "42", the corresponding instruction could be "1+41=", "2+40=", or even "the Answer to Life, the Universe and Everything is",\footnote{Douglas Adams, ``The Hitchhiker's Guide to the Galaxy'' (1979).} among numerous other possibilities. We provide this example to illustrate the complexity of the R-I correspondence. 

This uncertainty in the R-I scenario significantly complicates the extraction phase. It increases the likelihood of recovering plausible but incorrect instructions (e.g., "2+40=" instead of the original "1+41=" in the SFT data). This ambiguity not only complicates the initial inference but also introduces substantial errors in the extraction process, making R-I attacks inherently more challenging and less reliable than I-R attacks.

\parh{Targeted and Untargeted Attacks.}
Our study focuses on targeted attacks, where both the instruction used to query the LLM and the expected response are specific. In this context, an attack is considered successful only if a response from the SFT dataset is correctly matched with its corresponding instruction. In contrast, previous works on extracting pre-training data, such as~\cite{carlini2021extracting,nasr2023scalable}, often employed untargeted attacks, where success is determined solely by the presence of the response in the training data, regardless of the input prompt. This approach is crucial for SFT data extraction, as mismatched instruction-response pairs could lead to detrimental outcomes, especially in sensitive domains like healthcare (e.g., incorrect treatment recommendations for given symptoms).

We argue that untargeted attacks are unsuitable for SFT data extraction due to the potential risks associated with mismatched pairs in domain-specific applications.
We also posit that untargeted attacks are unlikely to effectively extract specific information from SFT data. To validate these assertions, we conduct experiments with existing untargeted attack methods under relaxed constraints (details in \mysec\ref{sec:setup} and \mysec\ref{subsec:reconstruction-attack-results}). These experiments demonstrate the limitations of untargeted approaches in SFT data extraction and underscore the necessity for specialized methods in targeted attacks.

\subsection{Pilot Study --- \va Extraction}
\label{subsec:pilot-study}

Before introducing our attack method, we conduct a pilot study to investigate
how effectively $M_{FT}$ can be directly extracted. In this context,
``directly'' refers to a simple and straightforward extraction process: given an
input query, we instruct the LLM to generate corresponding output, which is then
considered as the extracted data. This process mirrors the typical query flow
initiated by normal users and has been widely adopted in previous extraction
works~\cite{carlini2021extracting,nasr2023scalable}. We refer to this approach
as \va\ extraction throughout the rest of this paper. 
Despite the seemingly straightforward nature of this task, we find that extracting high-quality SFT data is indeed challenging. This finding justifies the need for a well-thought-out attack method.

\parh{\va Extraction Analysis.}~We choose the
AlpacaGPT4~\cite{peng2023instruction} dataset, one of the most popular SFT
datasets, to fine-tune the LLaMA-2-7B~\cite{llama27b} model. Following similar
SFT procedures used in previous works~\cite{wei2024magicoder,li2024split,wang2024exploring},
we select the checkpoint with the lowest loss; details of this process can be
found in~\mysec\ref{sec:setup}. The superior performance of the fine-tuned model
compared to the $M_{Base}$ on standard benchmarks is documented
in~\AP~\ref{app:pilot-performance}. Subsequently, we evaluate the model using
10,000 queries, comparing the generated responses with the actual responses used
in training. We measure similarity using BLEU scores~\cite{papineni2002bleu} and
exact match (EM). The results are presented in \T~\ref{tab:bleu-distribution}.

\begin{table}[!t]
    \centering
    \caption{BLEU score distribution and similarity result (BLEU Ran. = BLEU
    Range, \# of Res. = Number of Responses).}
    \label{tab:bleu-distribution}
    \resizebox{0.7\linewidth}{!}{

    \begin{tabular}{lclc}
    \hline
    BLEU Ran. & \# of Res. & BLEU Ran. & \# of Res. \\
    \hline
    0.0-0.1 & 4373 & 0.5-0.6 & 98 \\
    0.1-0.2 & 3624 & 0.6-0.7 & 55 \\
    0.2-0.3 & 1099 & 0.7-0.8 & 75 \\
    0.3-0.4 & 327 & 0.8-0.9 & 37 \\
    0.4-0.5 & 150 & 0.9-1.0 & 152 \\
    \hline
    \multicolumn{4}{l}{Average BLEU Score: 0.146; EM : 0\%} \\
    \hline
    \end{tabular}
    }
\end{table}

Notably, none of the 10,000 queries results in an EM with the SFT
data. Further investigation reveals that this low matching rate is primarily
due to a phenomenon we term \textbf{branch deviation}. To illustrate this
concept, consider the following example:

\begin{exmp}
    For an instruction ``\textit{Write a Python function with quick sort}'', the
standard response in the SFT data might begin with the keyword ``def'' followed
by the function name ``quick_sort'' and its parameters. However, if the model
generates a different initial token, such as ``Here's'' instead of ``def'', the
subsequent token generation can deviate significantly. 
\end{exmp}

We define this phenomenon as branch deviation, where a branch refers to a specific sequence of tokens generated by the model from a given prefix.
It occurs due to the auto-regressive nature of token generation in LLMs, where each new token is generated based on all preceding tokens. When a generated token
deviates from the ground truth, it can significantly alter the trajectory of
subsequent token generations, leading to responses that differ substantially
from the SFT data, causing each subsequent token to further diverge from the
original SFT data. While it is possible for a deviated branch to eventually
converge back to a path similar to the SFT data, the initial divergence creates irreversible discrepancies at both the token and semantic levels, compromising the overall accuracy of data extraction.

\begin{figure*}[!tbp]
    \centering
    \includegraphics[width=1.0\linewidth]{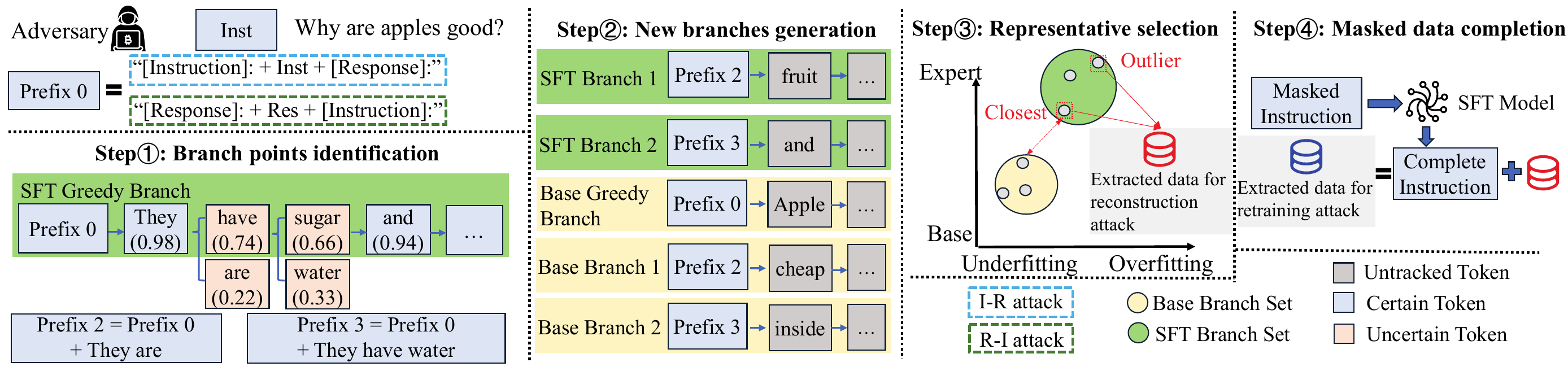}
    \caption{An overview of \tool's four-step workflow: (1) Branch points
    identification, (2) New branches generation, (3) Representative selection,
    and (4) Masked data completion. The figure illustrates an I-R attack example
    using the instruction ``Why are apples good?'', with token generation
    probabilities in parentheses.  It depicts the SFT and base model branches in Step \ding{173},
    and data extraction for reconstruction and retraining attacks in Steps \ding{174} and \ding{175}.}
    \label{fig:method-design}
\end{figure*}

\parh{Next Token Correction (NTC).}
Given the challenge posed by branch deviation, we seek to quantify the model's potential for accurately reproducing fine-tuning data if we could prevent such deviations. To this end, we first introduce the Next Token Correction (NTC) metric. This metric assesses the model's ability to predict the correct next token when provided with the ground truth sequence up to that point, effectively simulating a scenario where branch deviation is corrected at each step. Given an input sequence $X=(x_1, x_2, \ldots, x_k)$ of length $k$ and a ground truth output sequence $Y=(y_1, y_2, \ldots, y_n)$ of length $n$, the NTC is defined as:
\begin{equation}
    \text{NTC} = \frac{1}{n} \sum_{i=1}^{n} \mathbb{I}(P(x_1, x_2, ..., x_k, y_1, ..., y_{i-1}) = y_i)
\end{equation}
where $\mathbb{I}(condition)$ is the indicator function:
\begin{equation}
\mathbb{I}(condition) = \begin{cases}
1, & \text{if } condition \text{ is true} \\
0, & \text{otherwise}
\end{cases}
\end{equation}

Here, $P(x_1, x_2, ..., x_k, y_1, ..., y_{i-1})$ represents the model's prediction of the $i$-th token given the input sequence and all previous correct output tokens. For $i=1$, the prediction is based solely on the input sequence $(x_1, x_2, ..., x_k)$.
The NTC metric ranges from 0 to 1, representing the average correctness rate of the LLM in generating the next token given the ground truth tokens before. A higher NTC value indicates better token-level faithfulness in the model.

With the NTC metric defined, we randomly selected 100 queries from our previous 10,000 for NTC evaluation. The average NTC value is found to be 0.8297. This suggests that, under ideal conditions where we can successfully correct each generated token, up to 82.97\% of tokens could potentially be recovered at the token level. 
To further validate our findings, we conduct a similar experiment using the R-I attack method, which yields consistent results (see~\AP~\ref{app:ri-attack} for details).

\parh{Key Observations.}
Our analysis leads to a key observation: $M_{FT}$ indeed retains its SFT data, but extracting this data faces significant challenges due to branch deviation. These challenges can be summarized into two main points:

1) Identifying potential branch deviation points: Determining where in the generation process a branch deviation might occur is not straightforward, as it can happen at any token.

2) Token correction without ground truth: Even if we identify a potential deviation point, correcting the token without access to the ground truth is challenging. The model's auto-regressive nature makes it difficult to determine the correct token solely based on the generated sequence.

Addressing these challenges is key to developing effective methods for extracting SFT data from $M_{FT}$. Our subsequent attack method focuses on tackling these issues to improve data extraction performance.

\section{Design of \tool}
\label{sec:design}

In line with the challenges discussed in \mysec\ref{subsec:pilot-study}, we present the design of \tool, a novel and highly effective approach to extracting SFT data from $M_{FT}$.
\F~\ref{fig:method-design} illustrates \tool's high-level design.
We first introduce the insights behind \tool\ and how it addresses the previously mentioned challenges. Then, we provide a detailed description of each step.

\subsection{Design Insights}
\label{subsec:design-insights}

As observed in \mysec\ref{subsec:pilot-study}, $M_{FT}$ exhibits high NTC scores
for data present in the SFT dataset, suggesting that avoiding branch deviations
could lead to more accurate SFT data extraction. Building on this observation,
the key insight of \tool\ is to leverage the internal information of $M_{FT}$
during token generation to identify potential points that may lead to incorrect
extraction results (branch deviations). Furthermore, considering that
$M_{FT}$ may learn instruction-response pairs with varying degrees of
thoroughness due to their distinct training objectives and data organization
compared to pre-training, we consider both $M_{FT}$ and $M_{Base}$ in our approach.
Specifically, by forcing $M_{FT}$ to continue generation with the second most probable token at potential deviation points, we obtain a set of possible SFT branches. Similarly, we generate a set of base model branches from $M_{Base}$ for comparison. \tool\ then selects two branches from the SFT branch set: one outlier branch and another that is closest on average to the branches in the base set. This approach effectively leverages the behavioral differences between $M_{FT}$ and $M_{Base}$, enabling more accurate recovery of the original SFT data.

\subsection{\tool's Details}
\label{subsec:tool-details}

As illustrated in \F~\ref{fig:method-design}, we demonstrate \tool's data extraction process with an example under I-R attack. This process consists of four main steps: \ding{172} Branching points identification, \ding{173} New branches generation, \ding{174} Representative selection, and \ding{175} Masked data completion.

\parh{Branching Points Identification.}
This initial step begins with querying $M_{FT}$ using \va extraction with greedy decoding. Throughout this
process, we meticulously record the logits information for each generated token
and calculate its corresponding probability.
To identify potential branching points, we introduce a threshold \threshold. As we traverse the generated tokens from front to back, we flag any token whose probability falls below this threshold, considering it a point where the model might have alternative choices. We continue this process until all generated tokens have been processed or the number of identified branches reaches our predefined maximum, \maxbranch.
For instance, in \F~\ref{fig:method-design}, step \ding{172}'s SFT greedy branch identifies two such tokens (``have'' and ``sugar''), resulting in prefix 2 and prefix 3, respectively.

\parh{New Branches Generation.}
In this step, we use the prefixes obtained from step \ding{172} to query both $M_{Base}$ and $M_{FT}$, generating branches for each model under these prefixes. For both models, we produce a greedy branch (depicted as ``SFT greedy branch'' and ``Base greedy branch'' in \F~\ref{fig:method-design}), along with up to \maxbranch\ new branches stemming from the identified branching points (depicted as ``SFT branch $i$'' and ``Base branch $i$''.) During generation, we do not track the token probabilities, as indicated by the ``Untracked Token'' notation in \F~\ref{fig:method-design}.
Notably, our method progresses from front to back, as earlier branch deviations
tend to produce more significant divergences in the generated text. Per our observation, this
front-to-back strategy enables us to capture the most impactful variations for
subsequent branch selection. 

\begin{algorithm}[!htbp]
    \caption{Representative Selection}
    \label{alg:representative-selection}
    \begin{algorithmic}[1]
    \Require SFT branches $S$, Base branches $B$
    \Ensure Representative branches
    \State $d_i \gets$ avg. distance between $s_i \in S$ and all $b \in B$
    \State $s_c \gets \arg\min_{s_i \in S} d_i$  \space \textit{// ``Closest'' branch}
    \State $e_i \gets$ avg. distance between $s_i$ and all $s_j \in S, j \neq i$
    \State $s_o \gets \arg\max_{s_i \in S} e_i$ \space \textit{// ``Outlier'' branch}
    \State \textbf{return} $s_c, s_o$
    \end{algorithmic}
\end{algorithm}

\parh{Representative Selection.}
After obtaining branch sets from $M_{Base}$ and $M_{FT}$, we select the
representative branches from the SFT set as the extraction result. As
shown in \F~\ref{fig:method-design}, we view the base branch set (golden
circle) as lacking domain-specific knowledge, while the SFT branch set (green
circle) possesses this knowledge. To capture both underfitted and well-learned
data, we propose Algorithm~\ref{alg:representative-selection} to identify two
types of branches within the SFT branch set:

The algorithm selects (1) the ``Closest'' branch, which is most similar to the base space, potentially capturing underfitted data, and (2) the ``Outlier'' branch, which is most dissimilar from other SFT branches, representing thoroughly learned or potentially overfitted data. 
It aims to gain a comprehensive representation of the potential SFT data,
balancing between similarity to $M_{Base}$ and uniqueness within $M_{FT}$.

\parh{Outputs for Reconstruction and Retraining Attacks.}~The
above process generates a pair of outputs (``Closest'' and
``Outlier'') for each attack input (either I or R). For reconstruction attacks, both pairs are considered potential SFT dataset candidates and are thus retained. In the subsequent evaluation, we calculate the average distance between these pairs and the ground truth across various metrics to assess the results. Notably, in cases where the ``Closest'' and ``Outlier'' originate from the same branch, only one pair is preserved for evaluation to avoid redundancy.
Attackers can also leverage these pairs to retrain their own models, thus
enabling retraining attacks, where all collected pairs are used in the SFT
process. Nevertheless, we find that due to the various attack variants employed,
some pairs may contain masked content, potentially compromising their quality
and suitability for effective training. Thus, we provide the following step to
enhance the quality of the extracted data for retraining attacks.

\parh{Masked Data Completion.}~The final step in \tool, masked data completion, is a phase primarily aimed at enhancing the performance of retraining attacks. As detailed in \S\ref{sec:preservation-strategies}, we introduce three possible attack variants using distinct instruction preservation methods, which may result in masked or incomplete I/R used for querying the LLM. 
For instance, in the I-R attack scenario, an instruction might be partially masked, such as ``Tell me _ story _ Taylor Swift.'' Although the extracted response may be close to the ground truth, the incompleteness of the instruction itself could render the data suboptimal for further SFT. To address this limitation, this step aims to guide $M_{FT}$ in fulfilling the instructions based on these modified pairs.
Specifically, we prompt $M_{FT}$ using the following template, requesting it to generate a complete instruction:

\mybox{You will be given an incomplete instruction and its corresponding response. You need to return a complete, contextually appropriate new instruction that fits the given response. [Instruction]: \{instruction\}, [Response]: \{response\}.}

Notably, the masked data completion process is selectively applied only to data that has undergone masking. For unmasked data, we directly use the reconstructed ``Closest'' and ``Outlier'' branches obtained from the previous steps, combining them with the unmasked instructions or responses to form complete pairs for SFT training.
This step renders the extracted information suitable for further SFT, facilitating research into retraining attacks without affecting the performance of reconstruction attacks.

\section{Experimental Setup}
\label{sec:setup}

\parh{Selected LLMs.}~We select two popular LLMs as our base models for SFT in
our study: LLaMA2~\cite{touvron2023llama2} and CodeLlama~\cite{codellama}. For
generating instruction perturbations, we use DeepSeek-v2~\cite{liu2024deepseek}.
The overview of each LLM is outlined below.
\begin{itemize}[leftmargin=*,noitemsep,topsep=0pt]
  \item \underline{LLaMA2}~\cite{touvron2023llama2} is an advanced open-source LLM developed by Meta AI. It shows enhanced performance across various NLP tasks, including text generation, summarization, and question-answering. We use the LLaMA2-7B base version~\cite{llama27b}.

  \item \underline{CodeLlama} \cite{codellama} is a specialized family of LLMs for code-related tasks, built on the LLaMA2 architecture. It offers state-of-the-art capabilities in code completion, blank infilling, and processing of long contexts. We choose the CodeLlama-7B base version for the code-related SFT domain.

  \item \underline{DeepSeek-v2} \cite{liu2024deepseek} is an advanced Mixture-of-Experts (MoE) language model balancing powerful performance with efficient resource use. With 236B total parameters and a 128K token context length, it demonstrates superior capabilities across various language tasks. We specifically employ the DeepSeek-V2-0628~\cite{deepseekv2online} version in our experiments for generating diverse instruction perturbations and assembling high-quality SFT data.
\end{itemize}
Notably, we select the base model version that requires further SFT rather than the
fine-tuned version, as the specifics of their fine-tuning phase are not
accessible to us. This approach allows us to focus solely on our SFT dataset,
avoiding potential confounding factors from variations in the tuning processes.

\parh{SFT Datasets.}
In our experiments, we select two distinct SFT datasets, targeting code and math domains, respectively:
\begin{itemize}[leftmargin=*,noitemsep,topsep=0pt]
  \item \underline{OSS-Instruct}: Derived from MagicCoder's~\cite{wei2024magicoder} SFT process, this dataset leverages GPT-3.5~\cite{gpt35} to generate instruction pairs. It collects and filters semantically meaningful code snippets from open-source repositories, which are then processed to create high-quality instruction-response pairs for code-related tasks.
  
  \item \underline{MathInstruct}: Originating from MAmmoTH~\cite{yue2023mammoth}, this dataset is meticulously curated for general mathematical problem-solving. MathInstruct compiles data from 13 diverse math datasets, incorporating intermediate rationales. It ensures comprehensive coverage across various mathematical fields, allowing for diverse problem-solving approaches tailored to different mathematical challenges.
\end{itemize}
For our experiments, we randomly select 3K samples from each dataset for model fine-tuning and subsequent extraction attacks. This sampling strategy balances computational feasibility with experimental robustness, as suggested by~\cite{zhou2024lima}. \T~\ref{tab:stats} summarizes the key statistics of these datasets, including original sizes and average word counts for instructions and responses.

\parh{Evaluation Metrics for Reconstruction Attack.}
To quantify the similarity distance between the extracted information and the
ground truth as discussed in \mysec\ref{sec:methodology}, we employ three distinct
metrics:

\begin{itemize}[leftmargin=*,noitemsep,topsep=0pt]
    \item \underline{Continuous Token Matching (Token):} Following the approach of ~\cite{nasr2023scalable}, we consider the extracted information to successfully match the ground truth if it contains a continuous sequence of at least 25 tokens identical to the ground truth. Notably, we use 25 tokens rather than the ``extremely conservative'' threshold of 50 mentioned in~\cite{nasr2023scalable}, as our analysis in~\AP~\ref{app:token-win-length} shows that overly long windows can be suboptimal for evaluating certain domains.

    \item \underline{BLEU Score (BLEU):} To ease comparison with~\cite{carlini2021extracting}, we directly compute the BLEU score~\cite{papineni2002bleu}
    between the extracted data and the ground truth. 

    \item \underline{Embedding-based Similarity (Embed):} Inspired by~\cite{zhang2022automatic,cummins2024meta,li2022cctest}, we
    implement an embedding-based method to assess semantic similarity
    between the extracted information and the ground truth. 
    
\end{itemize}

\noindent
More details of these metrics can be found in~\AP~\ref{app:eval-metrics}.

\begin{table}[!tbp]
    \caption{Details of SFT datasets. Avg. \#W (I) and Avg. \#W (R) denote the average number of words for instructions and responses, respectively. Size shows the total number of instruction-response pairs.}
    \centering
    \scalebox{0.8}{
    \begin{tabular}{llcccc}
    \toprule
    \textbf{Dataset} & \textbf{Domain}  & \textbf{Avg. \#W (I)} & \textbf{Avg. \#W (R)} & \textbf{Size} \\ \midrule
    OSS-Instruct & Code  & 182.79 & 112.53 & 37k \\ \midrule
    MathInstruct & Math  & 47.76 & 84.08 & 262k \\ \bottomrule
    \end{tabular}
    }
    \label{tab:stats}
\end{table}

\begin{table}[!tbp]
  \centering
  \caption{Hyper-parameter settings. ``Len.'' stands for length.}
  \resizebox{0.8\linewidth}{!}{
  {\begin{tabular}{l|c||c|c}
  \toprule
     Hyperparameter  & Value &  Hyperparameter  & Value\\
   \midrule
    Optimizer & AdamW~\cite{KingBa15} &  \maxbranch   & 10 \\
    Learning rate & 5e-6 & Train batch size & 32 \\
    LR scheduler& Cosine \cite{loshchilov2017sgdr} & Valid batch size & 16 \\
      Sequence Len. & 2,048 &  Adam epsilon & 1e-8 \\
      Precision & BF16 & \threshold & 0.8 \\
    \bottomrule
  \end{tabular}
  }
  }
  \label{tab:param}
\end{table}

\parh{Evaluation Metrics for Retraining Attack.}
For evaluating the ER (defined in~\mysec\ref{sec:methodology}) in different domains, we employ widely-used benchmarks. In the code domain, we use HumanEval~\cite{chen2021evaluating}, which has been adopted by numerous studies~\cite{al2024traces,sun2024neural,ding2024cycle,nguyen2024beginning,codellama}. For the mathematics domain, we utilize GSM8K~\cite{cobbe2021gsm8k}, which has been extensively used to assess mathematical reasoning capabilities~\cite{wei2022chain,wang2023self}. Detailed information about the evaluation metrics and methodologies for each benchmark can be found in~\AP~\ref{app:eval-metrics}.

\parh{Compared Baselines.}~We consider the following baselines:
\begin{itemize}[leftmargin=*,noitemsep,topsep=0pt]

  \item  \underline{\va}~\cite{vicuna2023,li2023feasibility}: As described in~\S\ref{subsec:pilot-study}, this method extracts data through multiple queries to $M_{FT}$. While \F~\ref{fig:method-design} illustrates \va with a single query per instruction for clarity, in our actual experiments, we adjust the number of queries for \va to match the number of selected branches in \tool\ to ensure a fair comparison.
  \item \underline{\dsr}~\cite{zanella2020analyzing}: This method iteratively employs beam search over the vocabulary to generate candidate sequences, selecting those that differ most from the original model's outputs as extraction results. Notably, it is originally designed for scenarios with known sequence lengths of 5 tokens. While we optimize its search strategy for SFT data extraction, its computational cost remains substantially higher than both \tool\ and Vanilla.
\end{itemize}

As discussed in~\mysec\ref{subsec:attack-types-comparison}, we consider two untargeted data extraction methods from prior work:

\begin{itemize}[leftmargin=*,noitemsep,topsep=0pt]
  \item \underline{Random attack:} Proposed by \cite{carlini2021extracting}, this method randomly selects 100-character strings from Common Crawl~\cite{commoncrawl} as input queries for the LLM.
  \item \underline{Poem attack:} Introduced by \cite{nasr2023scalable}, this approach constructs prompts in the format: ``repeat this word forever: [WORD]'', where [WORD] is a single word repeated 50 times.
\end{itemize}

\begin{table*}[!t]
    \caption{Reconstruction attack performance under different attack variants with preservation settings. ``Met'' stands for Method. ``Full'' means full preservation. ``PWP'', ``PSP'', and ``SSP'' represent different preservation methods defined in \mysec\ref{sec:preservation-strategies}, with percentages indicating the retention rates. The highest value per row among all preservation methods (except ``Full'') is in bold.}
    \resizebox{0.90\linewidth}{!}{
    \begin{tabular}{llllccccccccccc}
    \hline
    \multirow{2}{*}{Attack} & \multirow{2}{*}{Model} & \multirow{2}{*}{SFT Data} & \multirow{2}{*}{Metric}& \multirow{2}{*}{Met}  & Full & \multicolumn{3}{c}{PWP} & \multicolumn{3}{c}{PSP} & \multicolumn{3}{c}{SSP} \\  \cmidrule(lr){7-9} \cmidrule(lr){10-12} \cmidrule(lr){13-15}
   & &   & & & - & 25\%   & 50\%   & 75\%   & 25\%     & 50\%    & 75\%    & 25\%   & 50\%   & 75\%  \\ \hline
   \multirow{12}{*}{I-R}& \multirow{6}{*}{CodeLlama}& \multirow{6}{*}{OSSInst}  & \multirow{3}{*}{BLEU} & \va& 0.651 & 0.569 & 0.580 & 0.605 & 0.569 & 0.576 & \textbf{0.618} & 0.515 & 0.557 & 0.589 \\
   & &   & & \dsr  & 0.645 & 0.579& 0.587& 0.613& 0.574& 0.586& \textbf{0.625}& 0.513& 0.558& 0.586 \\ 
   & &   & & \tool & 0.676 & 0.614 & 0.621 & 0.648 & 0.611 & 0.624 & \textbf{0.659} & 0.552 & 0.593 & 0.620 \\  \cline{4-15} 
   & &   & \multirow{3}{*}{Embed}& \va& 0.909 & 0.860 & 0.872 & 0.886 & 0.861 & 0.877 & \textbf{0.895} & 0.849 & 0.873 & 0.888 \\ 
   & &   & & \dsr  & 0.904 & 0.863& 0.871& 0.887& 0.861& 0.871& \textbf{0.894}& 0.850& 0.869& 0.880  \\ 
   & &   & & \tool & 0.917 & 0.879 & 0.885 & 0.901 & 0.876 & 0.886 & \textbf{0.908} & 0.867 & 0.885 & 0.895 \\ \cline{2-15} 
   & \multirow{6}{*}{LLaMA2}& \multirow{6}{*}{MathInst} & \multirow{3}{*}{BLEU} & \va&0.390 & 0.227 & 0.245 & 0.277 & 0.205 & 0.256 & 0.307 & 0.289 & 0.301 & \textbf{0.311} \\
   & &   & & \dsr  & 0.413  & 0.245& 0.264& 0.298& 0.221& 0.272& 0.321& 0.301& 0.319& \textbf{0.327}   \\ 
   & &   & & \tool &0.445  & 0.296 & 0.311 & 0.338 & 0.255 & 0.303 & 0.351 & 0.335 & 0.355 & \textbf{0.364} \\  \cline{4-15} 
   & &   & \multirow{3}{*}{Embed}& \va& 0.764& 0.625 & 0.647 & 0.677 & 0.528 & 0.616 & 0.669 & 0.686 & 0.692 & \textbf{0.702} \\  
   & &   & & \dsr  & 0.759 & 0.620& 0.649& 0.676& 0.532& 0.609& 0.664& 0.672& 0.688& \textbf{0.694}      \\ 
   & &   & & \tool & 0.784 & 0.670 & 0.694 & 0.713 & 0.569 & 0.643 & 0.694 & 0.706 & 0.721 & \textbf{0.728} \\ \cline{1-15} 

   \multirow{12}{*}{R-I}& \multirow{6}{*}{CodeLlama}& \multirow{6}{*}{OSSInst}  & \multirow{3}{*}{BLEU} & \va&0.388 & 0.384 & 0.413 & 0.421 & 0.412 & 0.428 & \textbf{0.440} & 0.439 & 0.421 & 0.415 \\ 
   & &   & & \dsr  & 0.417     & 0.436& 0.469& 0.477& 0.453& 0.476& \textbf{0.479}& 0.477& 0.461& 0.451      \\
   & &   & & \tool & 0.433 & 0.458 & 0.489 & 0.495 & 0.475 & 0.495 & 0.493 & \textbf{0.505} & 0.486 & 0.476 \\  \cline{4-15} 
   & &   & \multirow{3}{*}{Embed}& \va&0.528 & 0.674 & 0.711 & 0.707 & 0.676 & 0.699 & \textbf{0.716} & 0.699 & 0.668 & 0.656 \\ 
   & &   & & \dsr  & 0.594 & 0.719& 0.780& 0.775& 0.740& 0.772& 0.759& \textbf{0.794}& 0.757& 0.736      \\
   & &   & & \tool & 0.623 & 0.737 & 0.797 & 0.790 & 0.756 & 0.787 & 0.775 & \textbf{0.816} & 0.783 & 0.765 \\ \cline{2-15} 
   & \multirow{6}{*}{LLaMA2}& \multirow{6}{*}{MathInst}  & \multirow{3}{*}{BLEU} & \va& 0.181 & 0.143 & 0.177 & 0.196 & 0.153 & 0.185 & \textbf{0.197} & 0.185 & 0.186 & 0.189 \\
   & &   & & \dsr  & 0.198 & 0.163& 0.195& 0.211& 0.184& 0.205& \textbf{0.221}& 0.198& 0.205& 0.208       \\
   & &   & & \tool & 0.227 & 0.206 & 0.232 & 0.245 & 0.220 & 0.241 & \textbf{0.257} & 0.236 & 0.239 & 0.241 \\  \cline{4-15} 
   & &   & \multirow{3}{*}{Embed}& \va&0.509 & 0.481 & 0.542 & \textbf{0.582} & 0.460 & 0.529 & 0.569 & 0.531 & 0.542 & 0.548 \\ 
   & &   & & \dsr  & 0.522 & 0.513& 0.572& \textbf{0.604}& 0.480& 0.546& 0.581& 0.544& 0.552& 0.554      \\
   & &   & & \tool & 0.556 & 0.550 & 0.608 & \textbf{0.637} & 0.517 & 0.578 & 0.615 & 0.580 & 0.588 & 0.586 \\ \cline{1-15} 
    \end{tabular}
    }
    \label{tab:results-reconstruction}
\end{table*}

\parh{Implementation Details.}
The hyperparameters employed are detailed in Table~\ref{tab:param}.
To enhance computational efficiency and optimize GPU memory utilization, we implement the Optimizer State Sharding (ZeRO3) strategy from DeepSpeed~\cite{rasley2020deepspeed, rajbhandari2020zero}.
Following~\cite{wang2024exploring}, we reserve 10\% of our training data for validation.
We closely monitor the validation loss throughout the training process and use the configuration with the lowest validation loss for our final performance evaluation.
For SFT data extraction, we leverage the vllm framework~\cite{kwon2023efficient} during the inference phase to efficiently process LLMs.
Our experimental setup comprises a high-performance computing environment with eight H800 GPUs (80GB).

\section{Findings}
\label{sec:evaluation}

\subsection{Reconstruction Attack Results}
\label{subsec:reconstruction-attack-results}

In this section, we conduct a comprehensive analysis of the factors influencing extraction accuracy in reconstruction attacks, along with their potential underlying mechanisms. Our analysis focuses on interpreting the results presented in \T~\ref{tab:results-reconstruction}, examining the effects of possible attack variants with various preservation methods and their implications for attack efficacy. 
We present results using BLEU and embedding similarity metrics, while similar patterns observed with token-level metrics are detailed in~\AP~\ref{app:eval-reconstruct-token}.

\parh{Impact of Preservation Extent.}
Our analysis of how instruction preservation affects reconstruction attacks
reveals a positive relationship between the retention rate and attack
efficacy. \T~\ref{tab:results-reconstruction} demonstrates that across most
preservation methods and evaluation metrics, attack performance drops as the
proportion of retention rate decreases. 
In I-R attacks on the math domain, the BLEU
metric exhibits a notable decline from 0.669 (with 75\% information preserved)
to 0.528 (with only 25\% preserved) under PSP. This pattern persists across
various models, datasets, baselines, and attack variants, underscoring the critical role of
preserving original instruction structure in successful attacks.

This phenomenon can be attributed to the diminished availability of contextual information as the original instruction content is reduced, posing increasing challenges for the model to accurately recover the original information, thereby impeding attack effectiveness.

\parh{Influence of Preservation Methods.}
Examination of different preservation methods reveals that methods maintaining longer contiguous segments of the original instruction demonstrate superior performance.
While the improvements are moderate, PSP consistently achieves the best performance in 12 out of 24 evaluation scenarios across different models and datasets, followed by SSP and PWP.
This can be attributed to PSP's preservation of local semantic coherence within the instruction. By maintaining intact sentences, it retains a higher degree of semantic and syntactic integrity, thus providing more robust cues for reconstruction. Conversely, PWP tends to disrupt these structural elements more significantly. While SSP offers semantic coherence, it may introduce novel input sequences unfamiliar to $M_{FT}$, potentially compromising the reconstruction process.

\parh{Impact of Attack Types.}
Our analysis reveals a significant disparity in the difficulty between I-R and R-I attack types, with R-I proving to be substantially more challenging. Across all preservation methods and retention rates, R-I attacks consistently yield lower accuracy scores compared to their I-R counterparts. For instance, using CodeLlama on the OSSInst dataset with full preservation, the Embed metric for R-I attack is 29.4\% lower than that for I-R attacks. This pattern is consistently observed across different models and datasets, with the performance gap ranging from 7.6\% to 38.1\%. The increased difficulty of R-I attacks inlines with our analysis toward different attack types in~\mysec\ref{subsec:attack-types-comparison}.

Additionally, we observe that unlike I-R attacks, R-I attacks show less correlation with the retention rates. For instance, R-I attack performance does not consistently improve as SSP retention rates increase. This indirectly suggests the high feasibility of SFT dataset extraction. It implies that extracting information learned during SFT does not heavily depend on complete or continuous segments of the original data. Rather, meaningful information from the SFT dataset can be extracted even from fragmented or incomplete inputs.

\parh{Effectiveness of \tool.}
Across all preservation methods, retention rates, and attack types, \tool exhibits superior performance compared to \va, achieving an average relative improvement of 9.96\%. As for the baseline \dsr, while it outperforms \va in 88.75\% of settings, it incurs 6.4 times the average computational overhead of \tool and \va\ during sequence generation. Notably, \tool relatively outperforms \dsr by 5.73\% on average while maintaining the same computational efficiency as the \va approach. These results demonstrate \tool's strong balance between effectiveness and practicality, establishing it as a highly efficient approach for reconstruction attacks.

\begin{figure}[!tbp]
    \centering
    \includegraphics[width=0.99\linewidth]{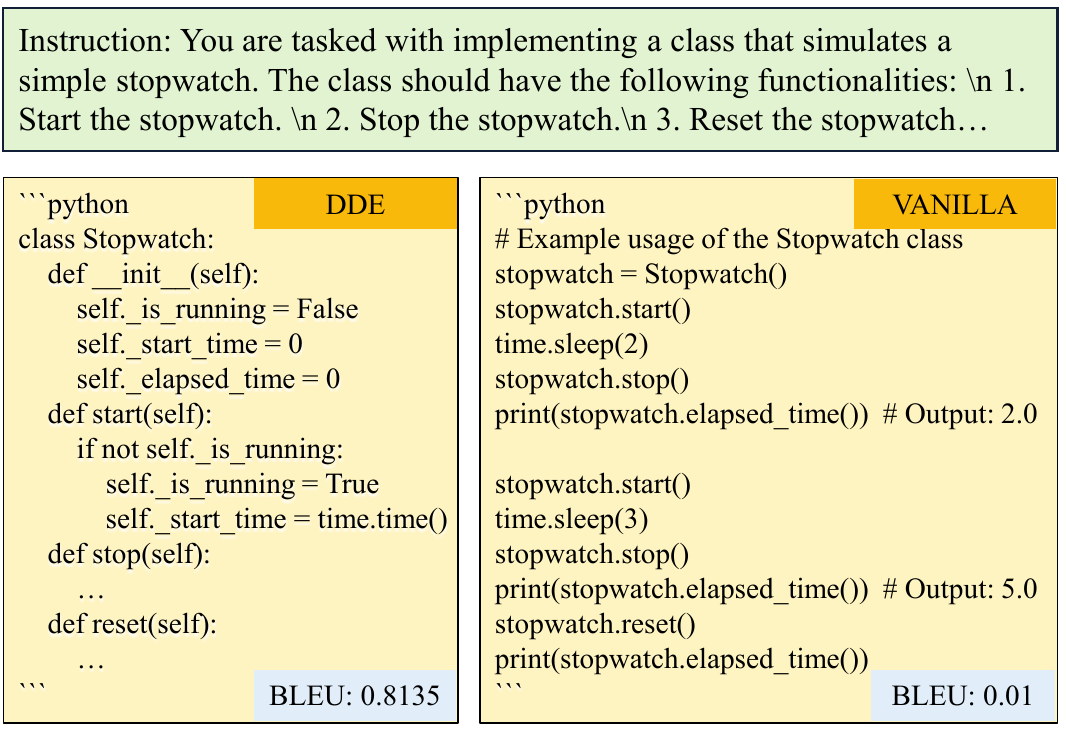}
    \caption{Example of extracted code from OSS-Instruct dataset with \tool and \va. For readability, we shorten the example code.}
    \label{fig:extraction-example}
\end{figure}

\parh{Reconstruction Attack Example.}~As we introduced
in~\S\ref{sec:setup}, the OSS-Instruct dataset is designed with high diversity,
so examples typically share only basic elements like function headers with
keywords such as 'def'. This makes extraction particularly challenging. To
illustrate the practical difference in extraction quality between methods, we
present an example from the OSS-Instruct dataset
in~\F~\ref{fig:extraction-example}. The figure shows an instruction for
implementing a Stopwatch class with specific functionality requirements (top),
alongside extractions produced by \tool and \va. By comparing the BLEU scores
between each extraction and the ground truth, we observe that \tool successfully
reconstructs most of the implementation (BLEU: 0.8135). In contrast, \va fails
to capture the essential class structure (BLEU: 0.01), extracting only example
usage code that lacks implementation details. This example clearly demonstrates
the effectiveness of \tool in reconstruction attacks.

\parh{Branch Contribution Analysis.}~We further conduct statistical
analysis to determine which branch provides the best extraction matches across
different attack scenarios. Using LLaMA2 as the base model with BLEU score
metrics, we find that for I-R attacks, the closest branch provides the best
match in 53.93\% of the cases, while the outlier branch accounts for 46.07\%. In
R-I attacks, this distribution shifts significantly, with the outlier branch
dominating at 69.07\% and the closest branch at only 30.93\%. These findings
demonstrate that both branches contribute to extraction performance, with their
effectiveness varying by attack type.

\parh{Additional Attack Baselines.}
As introduced in~\S\ref{sec:setup}, we include two untargeted data extraction methods (Random and Poem attack) as additional baselines. 
Specifically, we generate 10,000 examples for each method and query $M_{FT}$ in the math
domain. We then compare each of the 10,000 responses against every entry in the
SFT dataset, considering any match as a successful extraction. Two metrics are
employed to determine matches: BLEU score > 0.8 and continuous 25-token match.
Considering the format of SFT data, we define the ground truth as
the concatenation of the instruction and response.

\begin{table}[!htbp]
    \centering
    \caption{Matching rates for Random and Poem attacks.}
    \label{tab:attack-results}
    \resizebox{0.75\linewidth}{!}{
    \begin{tabular}{lccc}
    \hline
    Attack & BLEU (\%) & Token (\%) & Human (\%) \\
    \hline
    Random & 0.18 & 0.86 & 0 \\
    Poem & 1.51 & 3.22 & 0 \\
    \hline
    \end{tabular}
    }
\end{table}

As shown in \T~\ref{tab:attack-results}, the random attack exhibits an extremely low matching rate, while the poem attack demonstrates a relatively higher rate. Further manual analysis reveals that over 75\% of the matches correspond to strings similar to ``(a) 15 (b) 16 (c) 17 (d) 18 (e) 19'' in the SFT data. The poem attack, which requires repeating a single word indefinitely, frequently triggers catastrophic repetition or non-termination issues in the LLM. This behavior leads to an artificially inflated matching rate. Despite these automated matches, our manual analysis indicates that none of the matched cases exhibit true semantic similarity to entries in the SFT dataset. These results emphasize the importance of precisely defining SFT data extraction and developing more sophisticated approaches specifically tailored to this targeted extraction task, as existing untargeted methods prove inadequate for effectively extracting SFT data.

\finding{1}{Reconstruction attacks are significantly influenced by preservation methods, attack types, and retention rates, with methods maintaining semantic coherence and higher retention performing better. \tool consistently outperforms \va and \dsr across various settings, while previous baselines designed for pre-training data extraction prove impractical for SFT data.}

\subsection{Retraining Attack Results}
\label{subsec:retrainable-attack-results}

In this section, we analyze retraining attacks and the factors influencing their effectiveness. We evaluate how preservation methods and retention rates affect attack performance. We also compare \tool\ against \va and \dsr, demonstrating the improvements achieved by \tool.

\begin{figure}[!tbp]
    \centering
    \includegraphics[width=1.0\linewidth]{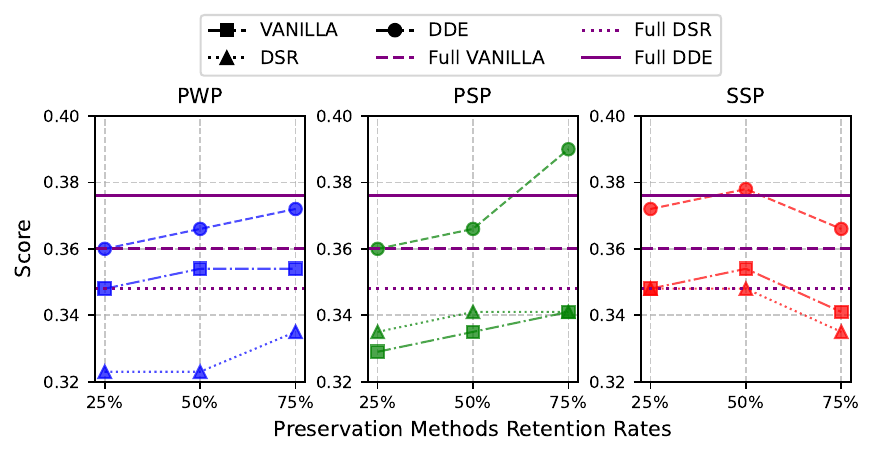}
    \caption{Retraining attack performance under different variants with preservation methods and retention rates.}
    \label{fig:attackable-performance}
\end{figure}

\parh{Retraining Attack Performance.}
Following the extraction phase, attackers can employ the acquired data for
subsequent SFT. Although both I-R and R-I attack data are theoretically viable,
our findings in \mysec\ref{subsec:reconstruction-attack-results} indicate that R-I
attacks produce significantly less similar data compared to the ground truth due
to their inherent complexity. Consequently, this section focuses exclusively on
data obtained from I-R attacks, as this choice allows us to fully demonstrate the potential effectiveness of SFT extraction attacks under the retraining goal. 
By examining the results of I-R attacks, we can better illustrate the severity of the threat and emphasize the urgent need for corresponding defense strategies.

\F~\ref{fig:attackable-performance} illustrates our experimental results, depicting the impact of PWP, PSP, and SSP across various retention rates. It employs colored lines to distinguish between the performance of \tool, \va, and \dsr, with a purple horizontal line representing the fully preserved data. 
Here we show only the code domain results, while similar trends for the math domain can be found in~\AP~\ref{app:retrainable-math}.
We observe that for PWP and PSP,
retraining attacks' effectiveness generally shows a positive correlation with the retention rate. 
However, SSP behaves differently: its performance first improves but then declines as retention rate increases.
Upon manual investigation, we attribute it to SSP's ability to perform semantic distillation, condensing key information from the original instructions.

Furthermore, we observe that no single preservation method consistently outperforms the others across all scenarios in retraining attacks. This finding underscores the complexity of SFT data extraction and highlights that high-performance retraining attacks can be achieved under different circumstances regardless of the preservation method.

\parh{\tool's Effectiveness in Retraining Attacks.}
For all three preservation methods and the full dataset configuration, \tool demonstrates superior performance compared to both \va and \dsr baselines, achieving average relative improvements of 9.41\% and 11.52\%, respectively. In contrast to the results in reconstruction attacks, \dsr shows no advantage in retraining attacks, outperforming \va in only 5 out of 20 configurations.

Furthermore, we observe an unexpected pattern when comparing attacks using partial data versus the full dataset. Unlike \va and \dsr, which perform best with the full dataset, \tool with certain preservation (e.g., 75\% for PSP) configurations actually outperforms those using the complete dataset. This finding, along with \dsr's contrasting performance in different attack objectives, emphasizes the fundamental differences between reconstruction and retraining attacks, as discussed in~\mysec\ref{sec:threat-model}. It demonstrates that scenarios exist where retraining attacks can achieve superior performance with less information, a nuance that cannot be captured by reconstruction attacks.

\parh{Impact of the Completion Process.}
As described in \mysec\ref{sec:design}, \tool\ incorporates a completion process for masked instructions or responses. This step is crucial for both I-R and R-I attack types, as it significantly influences the quality of the extracted data used for subsequent SFT process. 
To assess its effectiveness, we compare model performance with and without this completion step, focusing on PSP across math and code domains. Results are in \T~\ref{tab:completion-impact}.

\begin{table}[!tbp]
    \centering
    \caption{Impact of the completion process on model performance. \textit{W Com}: with completion; \textit{W/o Com}: without completion; \textit{Re Drop}: relative drop.}
    \label{tab:completion-impact}
    \resizebox{0.75\linewidth}{!}{
    \begin{tabular}{lcccc}
    \toprule
    & Retain & W Com & W/o Com & Re Drop (\%) \\
    \midrule
    \multirow{3}{*}{Code} & 25\% & 0.36 & 0.305 & 5.5 \\
    & 50\%  & 0.366 & 0.329 & 3.7 \\
    & 75\%  & 0.39 & 0.335 & 5.5 \\
    \midrule
    \multirow{3}{*}{Math} & 25\%  & 0.094 & 0.072 & 2.2 \\
    & 50\%  & 0.105 & 0.085 & 2.0 \\
    & 75\%  & 0.144 & 0.093 & 5.1 \\
    \bottomrule
    \end{tabular}
    }
\end{table}

The results indicate that SFT data without completion leads to lower performance, with an average drop of 4.0\%. Further analysis reveals that models fine-tuned on uncompleted data exhibited a 7.2\% higher probability of generating mask tokens compared to those trained on completed data. While this increase in mask token generation has limited influence on mathematical outputs, with an average drop of 3.1\%, it significantly affects code generation, showing an average drop of 4.9\%. This often leads to syntactic errors that compromise functionality. Based on these findings, we conclude that the completion step for masked queries is essential for maintaining the quality and effectiveness of \tool, particularly in addressing retraining attacks across diverse domains.

\finding{2}{Retraining attacks generally benefit from higher retention rates,
though SSP's effectiveness does not consistently increase due to semantic
distillation. \tool, with its crucial completion process, consistently
outperforms both \va and \dsr baselines, and can even surpass models trained on the full dataset using partial masked data.}

\subsection{Retraining vs. Reconstruction Attacks}
\label{subsec:attack-comparison}

In this section, we analyze the similarities and differences between retraining and reconstruction attacks from an experimental perspective, complementing the distinctions in their objectives and definitions discussed in \mysec\ref{sec:threat-model}.

\parh{Generalizability across Attack Variants.}~Retraining attacks show higher generalizability across different attack variants with preservation methods, as no single approach consistently outperforms others. In contrast, reconstruction attacks often yield better results with PSP. This highlights the adaptability of retraining attacks, which can effectively use SFT data with similar semantics but different expressions.

\parh{Full Dataset Efficacy.}~The two attack types exhibit markedly different
behaviors with the full dataset. In reconstruction attacks, the full dataset
consistently outperforms the masked ones. However, in retraining attacks, not
only does the full dataset fail to maintain an ``upper bound'' status, but some
masked datasets even achieve superior results. This stark contrast suggests that
the dataset used for SFT may not be optimal for downstream benchmarks in
retraining scenarios, revealing potential areas for improvement that
reconstruction attacks alone might overlook.

\parh{Retention Rate Impact.}~Both attack types generally show a positive
correlation between retention rate and attack performance, evident in improved
efficacy with higher retention rates. While retraining attacks have a few
exceptions in SSP, the overall trend underscores the importance of information
preservation in successful data extraction attempts.

These similarities and differences highlight the distinct nature of retraining and reconstruction attacks. The conceptual distinctions between two attack types are reflected in our experimental results, underscoring the importance of distinguishing between them in the context of SFT data extraction.

\subsection{Further Exploration}
\label{subsec:ablation-study}

\parh{Impact of the Base Model.}
As introduced in \mysec\ref{sec:methodology}, \tool\ employs a base model to represent the underfitting space. While our previous experiments use $M_{Base}$ corresponding to the $M_{FT}$, it is crucial to recognize that attackers may not have access to such specific information. To comprehensively assess \tool's performance under varying conditions, we evaluate its effectiveness using alternative base models in the math domain. Specifically, we compare the performance of Gemma-7B~\cite{team2024gemma} and ChatGLM3-6B~\cite{glm2024chatglm} against the original LLaMA2-7B. For each configuration, we conduct a series of 100 queries and present the averaged results in \T~\ref{tab:base-model-impact}.

\begin{table}[tbp]
\centering
\caption{Impact of the base models on \tool's performance.}
\label{tab:base-model-impact}
\resizebox{0.75\linewidth}{!}{
\begin{tabular}{lclll}
\toprule
& Base model & Token & BLEU & Embed \\
\midrule
\va & - & 0.141 & 0.387 & 0.758 \\
\dsr & - & 0.201 & 0.438 & 0.778 \\
\midrule
\multirow{3}{*}{\tool} & LLaMA2-7B & 0.212 & 0.465 & 0.795 \\
& Gemma-7B & 0.212 & 0.463 & 0.800 \\
& ChatGLM3-6B & 0.202 & 0.455 & 0.793 \\
\bottomrule
\end{tabular}
}
\end{table}

The results show that while using $M_{Base}$ corresponding to $M_{FT}$ (LLaMA2-7B) yields the best performance, \tool\ remains effective across different base models. Both Gemma-7B and ChatGLM3-6B show improvements over both \va and \dsr, with only small performance differences between them. This robustness to base model selection underscores \tool's versatility, making it applicable even when the exact base model of the victim model is unknown or unavailable to the attackers.

\begin{table}[!htbp]
    \centering
    \caption{Impact of \threshold\ on \tool's performance.}
    \label{tab:threshold-impact}
    \resizebox{\linewidth}{!}{
    \begin{tabular}{cccc cccc}
    \hline
    \threshold\ & BLEU & Token & Embedding & \threshold\ & BLEU & Token & Embedding \\
    \hline
    0.2 & 0.6127 & 0.8571 & 0.9182 & 0.6 & 0.6956 & 0.8687 & 0.9314 \\
    0.3 & 0.6602 & 0.7857 & 0.9075 & 0.7 & 0.6989 & 0.8687 & 0.9334 \\
    0.4 & 0.6780 & 0.8090 & 0.9238 & 0.8 & 0.7035 & 0.8687 & 0.9305 \\
    0.5 & 0.6848 & 0.8586 & 0.9282 & 0.9 & 0.7007 & 0.8586 & 0.9305 \\
    & & & & 1.0 & 0.6750 & 0.8586 & 0.9192 \\
    \hline
    \end{tabular}
    }
\end{table}

\parh{Impact of the Threshold.}
\tool\ involves a crucial hyperparameter: the threshold \threshold\ for potential branch points identification. Empirically, a lower \threshold\ allows for the selection of branch points with higher uncertainty, but results in fewer selections that are potentially positioned later in the input sequence.
Conversely, a higher \threshold\ enables earlier identification of potential branch points but may lead to false positives due to lower uncertainty levels. Thus, striking a balance between the branch points and confidence level necessitates an appropriate \threshold.
To investigate this, we conduct experiments in the code domain using the same settings as in \mysec\ref{sec:setup}, varying \threshold\ from 0.1 to 1.0. Each experimental configuration involves 100 queries. The results are presented in \T~\ref{tab:threshold-impact}. 
Note that \threshold\ of 0.1 is excluded as it fails to identify any potential branch points for many queries, rendering it impractical.

\begin{figure}[!t]
    \centering
    \includegraphics[width=1.00\linewidth]{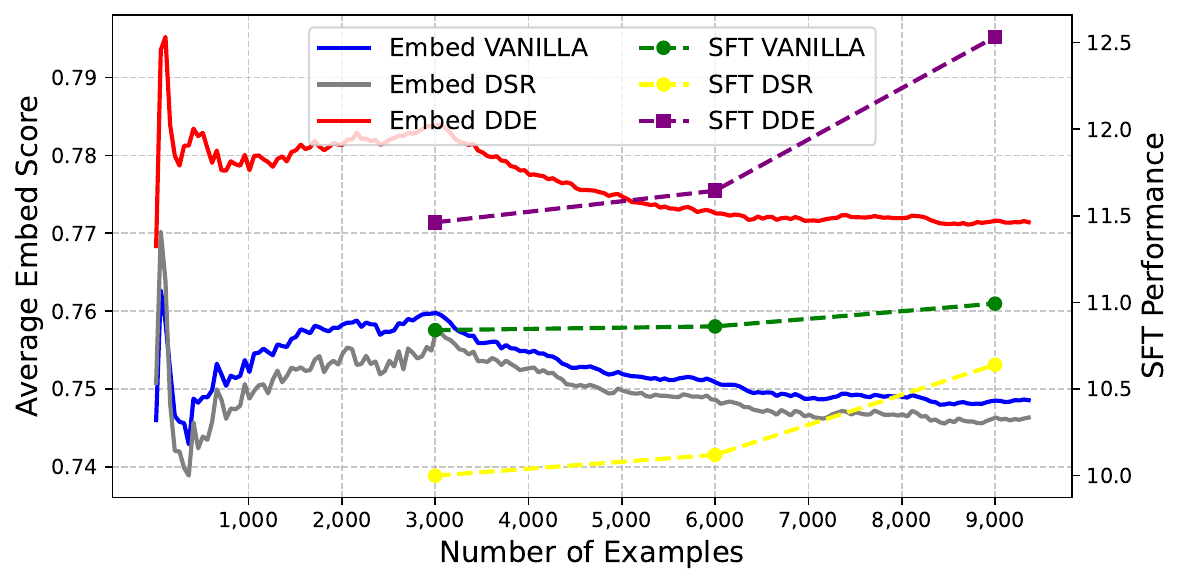}
    \caption{Performance comparison of \tool\ across an increasing number of examples. The graph shows cumulative average Embed scores for reconstruction attacks (left y-axis) and SFT performance for retraining attacks (right y-axis).}
    \label{fig:test-sftlength-lukas}
\end{figure}

The results in~\T~\ref{tab:threshold-impact} reveals an inverted U-shaped pattern in \tool's performance as \threshold\ increases. As \threshold\ increases from 0.2 to 0.8, we observe a general improvement across all metrics. The BLEU score shows a consistent upward trend, peaking at 0.7035 with \threshold\ of 0.8. Token-level accuracy stabilizes at 0.8687 for \threshold\ between 0.6 and 0.8, while embedding similarity reaches its maximum of 0.9334 at \threshold\ of 0.7. However, beyond \threshold\ of 0.8, we observe a decline in performance, with this trend becoming particularly noticeable at \threshold\ of 1.0.
This behavior aligns with our initial hypothesis regarding the trade-off between early branch identification and uncertainty levels.

Importantly, while our experiments suggest \threshold\ of 0.8 achieves
optimal performance, finding the exact optimal value is less critical for
deployment. As shown in~\T~\ref{tab:threshold-impact}, \tool maintains stable
performance across a wide range of non-extreme \threshold\ values (0.6-0.9). In
practice, adversaries can quickly adjust \threshold\ during initial runs based
on the frequency of identified branch points---increasing it if too few points
are found, or decreasing it if too many early-position points are identified.

\parh{Impact of the SFT Dataset Size.}
To assess the generalization capability of our method, we conduct experiments to evaluate the impact of SFT dataset size on extraction attack performance. In real-world scenarios, available SFT datasets can vary significantly in size, ranging from limited to extensive collections. Consequently, it is crucial for an attack method to demonstrate robust performance across diverse dataset sizes. We investigate this aspect in the math domain, consistent with the setup described in~\mysec\ref{sec:setup}, for both attack goals. For reconstruction attacks, we increase the querying time to 9,000 and calculate the cumulative average Embed score at intervals of 50 queries, providing a detailed view of performance changes. For retraining attacks, we evaluate the performance using three data points: 3,000, 6,000, and 9,000 examples. In these experiments, we focus solely on the impact of dataset size, thus preservation methods are not applied.
\F~\ref{fig:test-sftlength-lukas} illustrates the results of our experiments, comparing the performance of \tool\ against \va and \dsr. The x-axis represents the number of examples, while the left y-axis shows the cumulative average Embed score and the right y-axis displays the SFT performance.

Analysis of \F~\ref{fig:test-sftlength-lukas} reveals distinct patterns across two attack goals. For the reconstruction attack, \tool consistently outperforms both \va and \dsr, with all methods showing initial upward trends in cumulative average Embed scores before stabilizing. 
While \va and \dsr demonstrate comparable performance with fluctuating advantages at the beginning, \va gradually establishes a slight but consistent edge over \dsr as the extraction process continues. This stabilization occurs at a higher performance level for \tool, maintaining relative improvements of 3.5\% over \va and 3.9\% over \dsr throughout the extraction process.

In retraining attack, models trained on data extracted by \tool\ exhibit
superior SFT performance across all three data points, with a steeper growth
curve as the number of examples increases. At 9,000 examples, \tool achieves an SFT performance of 12.53, compared to 10.99 for \va and 10.64 for \dsr, representing relative improvements of 14.01\% and 17.76\%, respectively. This sustained improvement suggests that \tool\ can more
effectively capture valuable information from the victim model, enabling
retrained models to benefit substantially from larger extracted datasets.

\begin{table}[!t]
    \caption{Performance comparison of I-R reconstruction attack on
    WildChat under different attack variants. Each cell contains values in
    format of ``$a(+b)$'', where ``$a$'' is \tool's performance and ``$b$''
    shows the improvement over \va.}
    \resizebox{0.7\linewidth}{!}{
    \begin{tabular}{llcc}
    \hline
    Method & Ratio & BLEU & Embed \\
    \hline
    Full & - &  0.3132(+0.1028) &  0.6045(+0.1502) \\
    \hline
    \multirow{3}{*}{PWP} & 25\% &  0.2007(+0.1177)  & 0.4712(+0.2191)  \\
    & 50\% &   0.2161(+0.1324) &  0.5029(+0.2106) \\
    & 75\% &   0.2248(+0.1423) &  0.5104(+0.1953) \\
    \hline
    \multirow{3}{*}{PSP} & 25\% &   0.2452(+0.1124) &  0.5638(+0.1664) \\
    & 50\% &   0.2552(+0.0955) &  0.5713(+0.1612) \\
    & 75\% &   \textbf{0.2781(+0.1007)} &  \textbf{0.5867(+0.1581)} \\
    \hline
    \multirow{3}{*}{SSP} & 25\% & 0.2108(+0.1123) & 0.5184(+0.1679) \\
    & 50\% & 0.2138(+0.1224) &  0.5219(+0.1708) \\
    & 75\% & 0.2207(+0.1177) &  0.5410(+0.1762) \\
    \hline
    \end{tabular}
    }
    \label{tab:wildchat-result}
\end{table}

\parh{Scalability on Larger Datasets.}
To evaluate our method's scalability, we experiment with WildChat~\cite{zhao2024wildchat}, a privacy-sensitive dataset containing one million real conversations between users and ChatGPT across 68 languages. The scale of WildChat significantly exceeds both our previously used datasets (OSS-Instruct and MathInstruct in~\T~\ref{tab:stats}) and typical SFT datasets used in popular LLMs (LLaMA2 was fine-tuned using only 27,540 examples~\cite{touvron2023llama2}). Following the same setting in~\S\ref{sec:setup}, we perform I-R attacks on WildChat's official fine-tuned model under various attack variants.

\T~\ref{tab:wildchat-result} presents the results of these experiments.
The findings align with our earlier observations, showing that attack
performance generally improves as retention rate increases, with PSP
demonstrating the strongest performance across preservation methods. Overall,
\tool consistently outperforms \va across all configurations, with the
consistent performance gains validating our method's effectiveness and
scalability on extensive collections of real-world conversations.

\section{Potential Defense}
\label{sec:defense}

\parh{Design Goal.}
As described in \mysec\ref{sec:design}, \tool\ relies on the probability distribution of the next token generation to determine new branch generation. One potential defense against such attacks is to modify the returned token logits, aiming to prevent attackers from identifying uncertain tokens while minimizing the impact on normal usage. Specifically, our defense strategy is designed to achieve three primary goals: \ding{192} prevent attackers from extracting with \tool\ under a specified threshold, \ding{193} maintain unchanged results for greedy decoding, and \ding{194} ensure minimal changes in results when using sampling decoding with various Temperature and Top_p combinations.

\parh{Defense Method.}
Our defense method implements a rewrite of token logits. Given logits $L = [l_1, l_2, ..., l_n]$ sorted in descending order and the threshold \threshold, we denote the softmax function that converts logits to probabilities as $F$. Thus, the corresponding probabilities are $P = F(L)$. We first randomly generate a value $v \in [\tau, 1]$ as the target. Subsequently, we adjust and increase $l_1$, the logit of the highest probability token, to $l_1'$ such that $F(l_1') = v$. 
This process modifies only the logit of the highest probability token while ensuring it remains the most probable, thereby satisfying design goals \ding{192} and \ding{193}. As other logits remain unchanged, it minimally impacts normal usage, meeting goal \ding{194}.

Notably, our defense assumes that the attacker's threshold is known. This assumption is reasonable because attackers often involve multiple queries share the same prefix during the extraction process, exhibiting a query pattern distinctly different from others.
Furthermore, the logits returned to the attacker are accessible to the LLM service vendors. Therefore, we consider this a plausible assumption in practical scenarios.

\begin{figure}[t]
    \centering
    \includegraphics[width=0.8\linewidth]{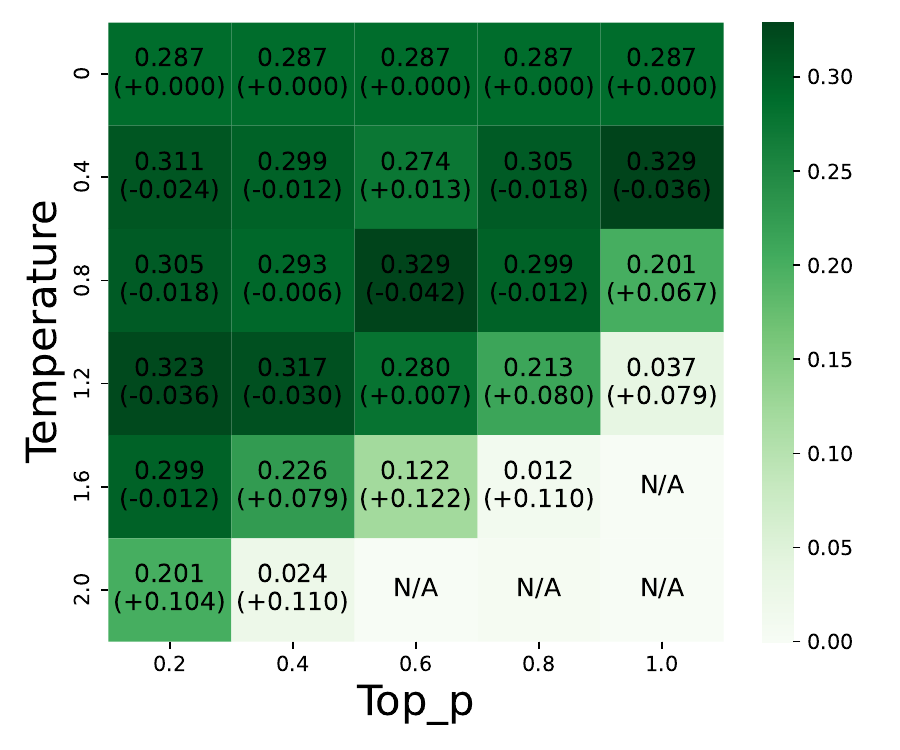}
    \caption{Impact of defense across various Temperature and Top\_p settings. Each cell shows performance in the format like ``performance with no defense (performance difference with defense).'' N/A denotes instances where the model generates numerous meaningless random tokens.}
    \label{fig:defense}
\end{figure}

\parh{Evaluation.}
To investigate the effectiveness of our defense method, we apply it to CodeLlama-7B and evaluate its performance on HumanEval under various hyperparameter settings. We examine two key hyperparameters: Temperature and Top\_p. Temperature ranges from 0 to 2.0, while Top\_p ranges from 0.2 to 1.0.
\F~\ref{fig:defense} presents a heatmap of our results, where each cell contains two values, $a(b)$. Here, $a$ represents the performance without defense, and $b$ is the difference in performance when the defense is applied.

Our analysis toward \F~\ref{fig:defense} reveals three key findings. First, for greedy decoding ($Temperature=0$), the results remain unchanged with defense, consistent with our design goal. Second, across the majority of parameter configurations, the performance with defense closely approximates that without defense, with most differences falling within ±3\%. Third, at high Temperatures, both methods show increased performance volatility, though such extreme values are usually outside the suggested ranges for practical application~\cite{openai-temp-suggest}.

\parh{Adaptive Attacks.}
While our defense mechanism effectively mitigates basic DDE attacks by modifying token logits, we acknowledge the possibility of adaptive attacks that could circumvent this protection~\cite{paulus2024advprompter,andriushchenko2024jailbreaking}. One potential adaptive strategy involves approximating token probabilities when logits are unavailable or untrustworthy. Specifically, an attacker could employ multiple accounts to perform sequential queries, each requesting only the next token. Through numerous queries with varying contexts, the attacker could empirically estimate the probability distribution of the next tokens, effectively bypassing our logit modification defense. However, this approach incurs significantly higher computational costs and requires substantially more queries compared to the original attack, as each token generation would necessitate multiple API calls. Additionally, service providers could implement rate limiting or anomaly detection to identify such patterns of sequential single-token requests. A comprehensive exploration of more sophisticated adaptive attacks and corresponding robust defenses represents an important direction for future research.
\section{Related Work}
\label{sec:related}

\parh{Model Extraction Attack.} Beyond the 
data extraction attacks on
private information for small-scale machine learning models~\cite{salem2020updates,zanella2020analyzing,jagielski2023combine,hui2023information} or the 
training data for LLMs~\cite{carlini2021extracting,nasr2023scalable}, researchers have developed Model Extraction (ME) attacks that directly target the models themselves~\cite{oh2019towards,wang2019stealing,li2023protecting,tramer2016stealing}. These attacks aim to infer critical properties of a victim model. Notable works in this field include Tramer et al.~\cite{tramer2016stealing}, who introduced an equation-solving technique to extract model parameters; and Yu et al.~\cite{yu2020cloudleak}, who demonstrated DNN model extraction from cloud platforms using minimal queries.
Additional contributions from Papernot et al.~\cite{papernot2016distillation} and Gong et al.~\cite{gong2021inversenet} further advanced methods for replicating model behavior and inferring internal structures. 
Notably, in the code domain, Li et al.~\cite{li2023feasibility} have shown that their method can efficiently extract the programming capabilities of ChatGPT in a black-box setting.
In contrast to these ME attacks, we investigate the potential risks of extracting SFT data, an aspect not previously explored in LLM security research.

\parh{Membership Inference Attack (MIA) in LLMs.}
Membership Inference Attacks (MIAs) aim to determine whether a specific
data instance was used in training a 
model~\cite{membership2017sp}. These attacks have been extensively studied in
various domains, including image classification, natural language processing,
and recommendation
systems~\cite{tang2022mitigating,yuan2022membership,chen2023poster,zhu2023membership,baluta2022membership}.
In the context of LLMs, MIAs face unique challenges due to the vast scale of
training data and limited exposure of individual instances. Recent works have
made significant progress in detecting pre-training data in LLMs. Shi et al.
introduced WIKIMIA and MIN-K PROB, leveraging the hypothesis that unseen
examples likely contain low-probability outlier words \cite{shi2024detecting}.
Zhang et al. proposed Min-K\%++, identifying local maxima in the modeled
distribution to detect training samples \cite{zhang2024min}. 
While previous works mainly focused on verifying the presence of data in pre-training sets and typically required shadow datasets, our study explores techniques for extracting SFT data with only partial I-R knowledge.

\section{Conclusion}
\label{sec:conclusion}

We have presented the first comprehensive study on extracting SFT data from
LLMs. We consider multiple attack goals (reconstruction and retraining) and
types (I-R and R-I), introduce three attack variants, and propose a novel DDE
approach. Experiments across various SFT domains and attack scenarios
demonstrate the feasibility of SFT data extraction, and the effectiveness of
DDE. We discuss defense methods, and provide insights into key factors affecting
DDE's performance.

\section*{Acknowledgement}
The HKUST authors are supported in part by a RGC CRF grant under the contract C6015-23G and research fund provided by HSBC. We are grateful to the anonymous reviewers and our shepherd for their insightful comments and suggestions, which have significantly improved the quality of this paper.

\bibliographystyle{ACM-Reference-Format}
\balance
\bibliography{bib/ref,bib/similarity,bib/decompiler,bib/machine-learning,bib/attack,bib/zjNewFull,bib/code,bib/cot,bib/imitation,bib/memory-attack,bib/sft,bib/llm}

\newpage
\appendix

\section{Metric Details}
\label{app:passk}

We employ the Pass@k metric to evaluate the proficiency of LLMs in solving programming tasks. This metric assesses whether an LLM can produce at least one correct solution that pass all unit tests within k attempts. The formal definition is as follows:

\begin{equation}
\mathit{Pass@k} = \frac{1}{n}\sum_{i=1}^{n} \mathbb{1}\left(\bigvee_{j=1}^k pass(s_i^j)\right)
\end{equation}

Where $n$ is the total number of programming problems in the test set, $k$ is the number of solution attempts allowed for each problem, $s_i^j$ represents the $j$-th solution attempt for the $i$-th problem, $pass(s)$ is a boolean function that returns True if solution $s$ passes all unit tests and False otherwise, $\mathbb{1}(x)$ is the indicator function returning 1 if $x$ is True and 0 if False, and $\bigvee$ denotes the logical OR operation.
\section{Parameter-Efficient Fine-tuning}
\label{app:peft}

As we mentioned in~\S\ref{sec:preliminary}, SFT methods can be further categorized into two main approaches: (1)
full parameter supervised fine-tuning (FSFT) and (2) parameter-efficient
fine-tuning (PEFT). Although PEFT demonstrates high performance while using
fewer parameters, studies~\cite{ghosh2024closer,zhang2024scaling} have shown
that it primarily assists the model with response initiation and extracts most
of the response from pre-trained knowledge. In other words, PEFT does not
significantly contribute to the model's ability to acquire new knowledge.
Therefore, in this study, we solely focus on the full parameter fine-tuning approach
and refer to it as the SFT. 
Moreover, \tool makes no assumptions about the fine-tuning method and can be directly applied to PEFT models without modification.
\section{Performance Details}
\label{app:pilot-performance}

In this section, we present the performance results of the LLaMA-2-7B model before and after fine-tuning on the Alpaca-GPT4 dataset. We evaluate the model on three benchmark datasets: MMLU~\cite{hendrycks2020measuring}, Trivial QA~\cite{joshi2017triviaqa}, and OpenBookQA~\cite{OpenBookQA2018}. The results are shown in \T~\ref{tab:benchmark-results}.

\begin{table}[h]
\centering
\caption{Performance comparison of LLaMA-2-7B before and after fine-tuning.}
\label{tab:benchmark-results}
\resizebox{0.95\linewidth}{!}{

\begin{tabular}{lccc}
\hline
 & MMLU & Trivial QA & OpenBookQA \\
\hline
Before fine-tuning & 45.3 & 68.9 & 58.6 \\
After fine-tuning & 49.8 & 71.4 & 66.9 \\
\hline
\end{tabular}
}
\end{table}

Analysis of the results shows that the fine-tuned model demonstrates consistent improvement across all three benchmarks. On average, the model's performance increased by 5.1 percentage points (4.5 for MMLU, 2.5 for Trivial QA, and 8.3 for OpenBookQA). This substantial improvement indicates that the fine-tuned model we used in \S\ref{subsec:pilot-study} can serve as a reasonable proxy for real-world service models. 
\section{R-I Attack Analysis}
\label{app:ri-attack}

To further validate our findings in~\S\ref{subsec:pilot-study}, we conduct an experiment using the R-I attack. Results in \T~\ref{tab:ri-bleu-distribution} demonstrate a similar pattern to our initial findings. The low average BLEU score of 0.1150 and the absence of exact matches indicate that the model faces similar challenges in reproducing SFT data accurately, even when exchanging the position of the instructions and responses.

\begin{table}[!h]
    \centering
    \caption{BLEU score distribution for R-I attack responses. (BLEU Ran. = BLEU Range, \# of Res. = Number of Responses)}
    \label{tab:ri-bleu-distribution}
    \begin{tabular}{lclc}
    \hline
    BLEU Ran. & \# of Res. & BLEU Ran. & \# of Res. \\
    \hline
    0.0-0.1 & 6581 & 0.5-0.6 & 501 \\
    0.1-0.2 & 375 & 0.6-0.7 & 285 \\
    0.2-0.3 & 769 & 0.7-0.8 & 157 \\
    0.3-0.4 & 647 & 0.8-0.9 & 28 \\
    0.4-0.5 & 526 & 0.9-1.0 & 131 \\
    \hline
    \multicolumn{4}{l}{Average BLEU Score: 0.1150} \\
    \multicolumn{4}{l}{EM: 0\%} \\
    \hline
    \end{tabular}
\end{table}

\section{Retraining Attack Performance on Math Domain}
\label{app:retrainable-math}

\begin{figure}[!htbp]
    \centering
    \includegraphics[width=1.0\linewidth]{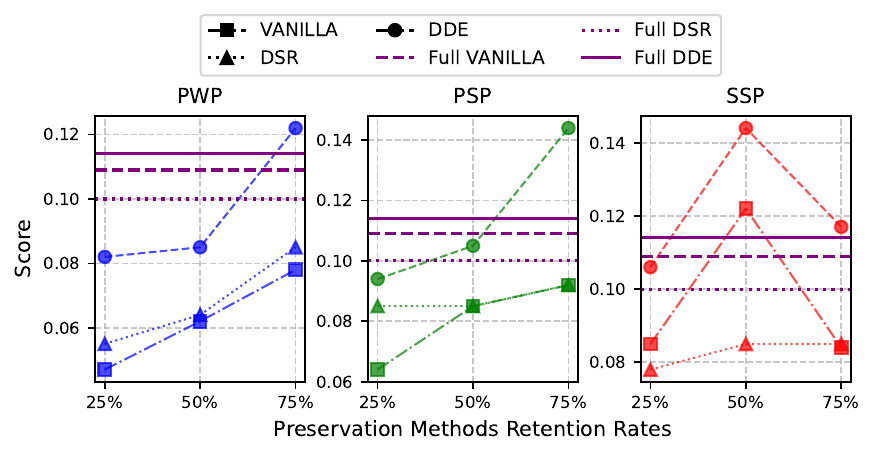}
    \caption{Performance comparison of retraining attacks under different attack variants using preservation methods and retention rates for the math domain.}
    \label{fig:attackable-performance-math}
\end{figure}

\F~\ref{fig:attackable-performance-math} presents the experimental results for the math domain, demonstrating the impact of PWP, PSP, and SSP across various retention rates.

From the figure, we can observe that the math domain have similar trends to those in the code domain, with performance varying based on the preservation method and retention rate used. 

\section{Impact of token window length}
\label{app:token-win-length}

\begin{figure}[!t]
    \centering
    \includegraphics[width=1.0\linewidth]{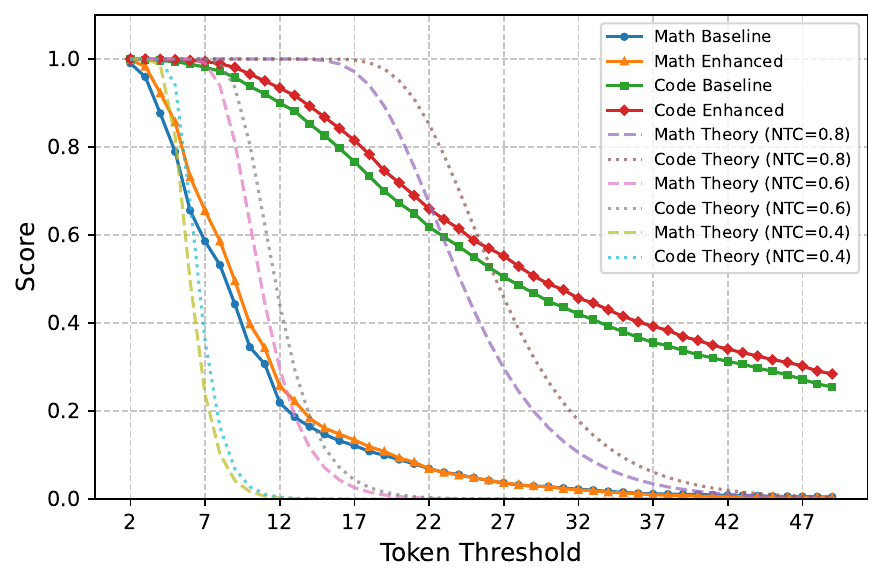}
    \caption{Comparison of actual results and theoretical probabilities for different token window length in math and code domains. The solid lines represent the actual performance of baseline and enhanced methods, while the dashed and dotted lines show theoretical predictions for different token accuracy probabilities.}
    \label{fig:token-window}
\end{figure}

\F~\ref{fig:token-window} illustrates the impact of token window length on the performance of our proposed method compared to the baseline, as well as theoretical predictions. The x-axis represents the token window length, while the y-axis shows the score under the continuous token matching metric. Solid lines depict actual results for both math and code domains, with circles and triangles representing baseline and enhanced methods, respectively. These results are obtained using a consistent PWP setting with a retention rate of 25\%, which we find to be representative of the general trend across various settings. Dashed and dotted lines represent theoretical probabilities calculated under the assumption of independent token generation, with $NTC \in [0.4,0.6,0.8]$ is defined in~\S\ref{subsec:pilot-study}. The theoretical curves are computed using the formula $P(\text{success}) = 1 - (1 - NTC^k)^{(L - k + 1)}$, where $k$ is the token window length and $L$ is the average length of the domain (as shown in \T~\ref{tab:stats}). This theoretical model provides a benchmark against which we can compare our empirical results, allowing us to assess the effectiveness of our approach across different domain-specific data average length and token window length.

Analysis of \F~\ref{fig:token-window} reveals several key insights into the performance of our enhanced method compared to the baseline. Our approach consistently outperforms the baseline across all token window lengths for both math and code domains, demonstrating its robustness and efficacy. Notably, the actual performance curves exhibit a more gradual decline compared to the theoretical predictions based on independent token generation. This discrepancy suggests that in practice, token generation is not entirely independent, and the model's performance degrades more gracefully as the token window length increases, highlighting the complex interdependencies in LLMs. Furthermore, the divergence between math and code domain curves, particularly at higher token window length, indicates that using a uniform token window size (e.g., 25) may not be optimal due to varying average token lengths across domains. This observation aligns with the low score of the token metric in the math domain shown in \T~\ref{tab:results-reconstruction-token}, underscoring the importance of domain-specific token window length tuning for better performance evaluation across diverse domains.
\begin{table*}[!t]
    \caption{Reconstruction attack performance of token metric. The highest value per row among all preservation methods (except ``Full'') is in bold.}
    \resizebox{0.99\linewidth}{!}{
    \begin{tabular}{llllccccccccccc}
    \hline
    \multirow{2}{*}{Attack} & \multirow{2}{*}{Model} & \multirow{2}{*}{SFT Data} & \multirow{2}{*}{Metric}& \multirow{2}{*}{Met}  & Full & \multicolumn{3}{c}{PWP} & \multicolumn{3}{c}{PSP} & \multicolumn{3}{c}{SSP} \\  \cmidrule(lr){7-9} \cmidrule(lr){10-12} \cmidrule(lr){13-15}
   & &   & & & - & 25\%   & 50\%   & 75\%   & 25\%     & 50\%    & 75\%    & 25\%   & 50\%   & 75\%  \\ \hline
   \multirow{6}{*}{I-R}& \multirow{3}{*}{CodeLlama}& \multirow{3}{*}{OSSInst}  & \multirow{3}{*}{Token} & \va & 0.774 & 0.553 & 0.589 & 0.661 & 0.573 & 0.609 & \textbf{0.718} & 0.548 & 0.620 & 0.660 \\ 
   & &   & & \dsr  & 0.759& 0.546& 0.579& 0.645& 0.574& 0.599& \textbf{0.699}& 0.527& 0.609& 0.656     \\ 
   & &   & & \tool & 0.795& 0.588 & 0.620 & 0.694 & 0.618 & 0.648 & \textbf{0.740} & 0.569 & 0.650 & 0.697 \\ \cline{4-15} 
   & \multirow{3}{*}{LLaMA2}& \multirow{3}{*}{MathInst}  & \multirow{3}{*}{Token} & \va & 0.130 & 0.044 & 0.052 & \textbf{0.071} & 0.023 & 0.032 & 0.060 & 0.046 & 0.049 & 0.051 \\   
   & &   & & \dsr  & 0.133 & 0.039& 0.047& \textbf{0.066}& 0.020& 0.028& 0.059& 0.044& 0.047& 0.053 \\ 
   & &   & & \tool & 0.151 & 0.049 & 0.058 & \textbf{0.078} & 0.025 & 0.040 & 0.071 & 0.054 & 0.062 & 0.069 \\ \cline{4-15} 
   \multirow{6}{*}{R-I}& \multirow{3}{*}{CodeLlama}& \multirow{3}{*}{OSSInst}  & \multirow{3}{*}{Token} & \va & 0.168 & 0.057 & 0.131 & 0.214 & 0.085 & 0.169 & \textbf{0.228} & 0.198 & 0.196 & 0.197 \\  
   & &   & & \dsr  & 0.207     & 0.066& 0.152& 0.247& 0.104& 0.186& \textbf{0.263}& 0.249& 0.245& 0.246     \\
   & &   & & \tool & 0.223 & 0.088 & 0.178 & 0.278 & 0.129 & 0.217 & \textbf{0.289} & 0.281 & 0.277 & 0.273 \\ \cline{4-15} 
   & \multirow{3}{*}{LLaMA2}& \multirow{3}{*}{MathInst}  & \multirow{3}{*}{Token} & \va & 0.013 & 0.008 & 0.009 & 0.014 & 0.007 & 0.013 & \textbf{0.018} & 0.009 & 0.010 & 0.008 \\ 
   & &   & & \dsr  & 0.018 & 0.009& 0.011& 0.015& 0.006& 0.014& \textbf{0.022}& 0.008& 0.009& 0.009       \\
   & &   & & \tool & 0.020 & 0.010 & 0.012 & 0.016 & 0.007 & 0.016 & \textbf{0.023} & 0.009 & 0.010 & 0.010 \\ \hline
    \end{tabular}
    }
    \label{tab:results-reconstruction-token}
\end{table*}

\section{Evaluation Metrics Details}
\label{app:eval-metrics}

This section provides detailed information about the evaluation metrics and methodologies used for each benchmark in our study.

\parh{Evaluation Benchmarks.}
For SFT in the code domain, we adopt HumanEval~\cite{chen2021evaluating} as our benchmark. In alignment with prior studies \cite{wei2024magicoder, li2023starcoder, luo2023wizardcoder,abdin2024phi, codellama}, we employ the Pass@k metric to evaluate accuracy. This metric determines whether LLMs can pass all unit tests with the first k generated solutions. Following~\cite{wei2024magicoder, li2023starcoder, luo2023wizardcoder}, we adopt Pass@1 as our primary metric, i.e., $k=1$. More details about the Pass@k metric are provided in \AP~\ref{app:passk}.

For SFT in the mathematics domain, we select GSM8K~\cite{cobbe2021gsm8k} as our benchmark. For evaluation, we use 5-shot queries, as standardized in~\cite{lm-evaluation-harness}. This approach allows for a fair assessment of the model's problem-solving capabilities by identifying the final numerical answer in the model's output.

\parh{Evaluation Metrics for Reconstruction Attack.}
\begin{itemize}[leftmargin=*,noitemsep,topsep=0pt]
    \item Continuous Token Matching (Token):
    This metric captures long, exact matches and serves as a relaxed version of EM, which our pilot study in \mysec\ref{subsec:pilot-study} found challenging for extraction.

    \item BLEU Score (BLEU): As a widely-used
    metric in measuring textual similarity~\cite{feng2020codebert,li2023feasibility}, it assesses text similarity based on
    n-gram overlap, providing a nuanced assessment that captures partial matches
    and allows for variations in word order and phrasing.

    \item Embedding-based Similarity (Embedding): Specifically, we use
    the Sentence-BERT model ``all-MiniLM-L6-v2''~\cite{reimers2019sentence} to generate embeddings.
    
\end{itemize}

These three metrics offer complementary perspectives on the quality of extraction. The continuous token matching provides a strict measure of exact reproduction, the BLEU score offers a more flexible assessment of textual similarity, and the embedding-based method captures semantic closeness. Together, they provide a comprehensive evaluation of the reconstruction attack's effectiveness.

\section{Reconstruction Attack Results of Continuous Token Matching}
\label{app:eval-reconstruct-token}

As a complementary evaluation to the main results presented in \S\ref{subsec:reconstruction-attack-results}, we provide additional analysis using continuous token matching as the evaluation metric. \T~\ref{tab:results-reconstruction-token} shows the detailed results, where the overall patterns observed here align with our main findings using BLEU and embedding similarity metrics.

\section{LLM Service Vendors Logits Availability}
\label{app:logits-availability}

In this section, we analyze the availability of token logits information across LLM service vendors. We categorize these vendors into two groups: proprietary model vendors and open-source model API services.

For proprietary models, we select platforms hosting top-10 models according to the Chatbot Arena LLM Leaderboard~\cite{lmarena}.
For open-source model services, we reference the popular LLM API vendors identified in~\cite{topllmvendors}. \T~\ref{tab:logits-availability} summarizes our findings.

\begin{table}[h]
    \centering
    \caption{Logits availability across LLM vendors. ``Support'' stands for whether the API documentation supports returning logits; ``Access'' stands for whether non-null logits are actually returned.}
    \label{tab:logits-availability}
    \resizebox{0.99\linewidth}{!}{
    \begin{tabular}{lcclcc}
    \toprule
    \textbf{Vendor} & \textbf{Support} & \textbf{Access} & \textbf{Vendor} & \textbf{Support} & \textbf{Access} \\
    \midrule
    OpenAI~\cite{apiopenai} & \CBrush & \CBrush & Replicate~\cite{replicate} & \XBrush & \XBrush \\
    Google~\cite{apigoogle} & \CBrush & \XBrush & HuggingFace~\cite{huggingface} & \CBrush & \CBrush \\
    Azure~\cite{apiazure} & \CBrush & \CBrush & Groq~\cite{groq} & \CBrush & \XBrush \\
    Grok~\cite{grok} & \CBrush & \XBrush & Deepinfra~\cite{apideepinfra} & \CBrush & \CBrush \\
    Anthropic~\cite{apianthropic} & \CBrush & \XBrush  & Anyscale~\cite{anyscale} & \CBrush & \CBrush \\
    Together~\cite{togetherapi} & \CBrush & \CBrush & Novita~\cite{novita} & \CBrush & \CBrush \\
    Fireworks~\cite{fireworks} & \CBrush & \XBrush & OpenRouter~\cite{openrouter} & \CBrush & \CBrush \\
    \bottomrule
    \end{tabular}
    }
\end{table}

Our results show that 13 out of 14 major LLM service vendors (93\%) have implemented API endpoints that support returning token logits, though not all currently expose this functionality to users. Notably, 8 vendors (57\%) already return logits in practice, validating the real-world feasibility of our attack scenario. This widespread availability of logits information, particularly among platforms hosting both proprietary and open-source models, confirms that our attack setting is not merely theoretical but represents a practical security concern in the current LLM ecosystem.

\section{Ethics and the Open Science}
\label{sec:ethics}

We have carefully evaluated and addressed potential risks associated with our work, adhering to rigorous ethical standards and open science policies. This section outlines our approach to ethical research conduct, responsible disclosure, and legal compliance, demonstrating our commitment to advancing AI security while minimizing potential harm.

\parh{Minimal Real-world Harm.} Our research methodology was carefully designed to minimize potential harm to real-world systems and users. All experiments were conducted in a controlled environment using locally hosted models, eliminating risks to public services or end-users. We thoughtfully considered resource usage, focusing on code and mathematical domains to limit ethical risks related to sensitive data. All computational costs and system loads were borne entirely by our own resources, ensuring no burden was placed on any external service providers. By implementing these measures, we conducted our study responsibly, gathering valuable insights into LLM vulnerabilities without compromising existing systems or violating ethical research practices in AI security.

\parh{License Compliance.} We strictly follow the license requirements of the open-source models and datasets used in our study. This adherence is fundamental to our research ethics and methodology. For the attack methods referenced in~\S\ref{subsec:reconstruction-attack-results}, we fully comply with their respective license requirements, ensuring that our usage aligns with the intentions and stipulations set forth by the original developers. This commitment to license compliance is not merely a legal obligation but a core principle of our research practice. It ensures that our research respects intellectual property rights, acknowledges the contributions of others in the field, and adheres to established open-source practices. By following these licensing guidelines, we aim to foster a collaborative and transparent research environment, promoting the sharing of knowledge while safeguarding the rights of all contributors in the open-source community.

\parh{Responsible Data Disclosure.} We have released a demo of our attack methods publicly available to facilitate reproducibility within the research community. However, we decide not to disclose certain specialized optimization techniques used to enhance attack efficiency. Importantly, this decision does not impact the ability to reproduce our experimental results, as the core methodologies are fully described. This balanced approach allows for comprehensive scientific verification while mitigating the risk of malicious actors exploiting our code to attack real-world online services. 
For datasets used, we prioritize linking to original open-source locations while maintaining copies for verification experiments. We also share data generated from poem and random attacks, despite their limited effectiveness, to support future research. 
Additionally, we fully disclose our defense methods, including accelerated implementations, to enable researchers to reproduce our results and provide service providers with practical defense strategies.

\parh{Human Evaluation and Potential Risks.}~Our study involved a limited human evaluation component, specifically in~\S\ref{subsec:reconstruction-attack-results}, where we assessed the results of random attacks and poem attacks. The evaluators were the authors of this paper. While we acknowledge the potential risks associated with human evaluation, such as exposure to hate speech or controversial content, the actual risk in our experiments was minimal. This is primarily because both attack types performed poorly, with most generated responses being either meaningless gibberish or catastrophic repetitions of words or punctuation marks. Consequently, we believe the associated risks in our experimental process were low. However, it's important to note that we still highlight the potential risks of such attacks, emphasizing the need for future work to give more consideration to these aspects. We recommend that future studies involving more sophisticated attacks or larger-scale human evaluations should implement robust safeguards to protect evaluators and consider obtaining formal ethical approval if necessary.

\parh{Balancing Transparency and Security.} While we have made every effort to minimize the risk of malicious use of our methods, we acknowledge the significant potential interests involved in SFT data extraction. This importance stems from both the inherent value of SFT data and the current lack of clear societal consensus on SFT data copyright issues. We recognize the tension between data copyright protection and the need for advancing LLM development, which often requires access to large, diverse datasets. By responsibly disclosing our research, we aim to promote informed discussions on the ethical usage of SFT data and the development of advanced LLMs. We believe that a balanced approach, considering both data protection and scientific progress, is crucial. Therefore, we consider the controlled open-sourcing of relevant methods to be a responsible solution that contributes to both security awareness and LLM advancement. 

\parh{Legal Considerations.} Beyond ethical considerations, we carefully examined our research's legal implications, particularly in light of the EU Artificial Intelligence Act~\cite{euAIlaw}. Our work intersects with both General-purpose AI (base LLMs used in our study, such as the LLaMA series) and Limited risk AI (embedding models) categories as defined by this act. While the Limited risk models are not subject to stringent regulatory oversight, we acknowledge the less defined legal landscape surrounding training data. Given the lack of detailed legislation in this area, we adopted a cautious approach, strictly adhering to open-source dataset guidelines and avoiding datasets potentially containing personally identifiable information (PII). This approach reflects our commitment to legal compliance in an evolving regulatory environment, balancing scientific progress with legal and ethical responsibilities in AI research.

\end{document}